\documentclass[prl,twocolumn,superscriptaddress,showpacs,aps,amssymb,floatfix]{revtex4-1}
\usepackage{amsmath,amssymb,bm}
\usepackage{graphicx}
\usepackage{epstopdf}
\usepackage{latexsym}
\usepackage{subfigure}
\usepackage{color}
\usepackage{multirow}
\usepackage{hhline}
\usepackage{comment}
\usepackage{tabularx}
\usepackage{tikz}

\begin{document}

\title{Spinon magnetic resonance of quantum spin liquids}
\author{Zhu-Xi Luo}
\author{Ethan Lake}
\affiliation{Department of Physics and Astronomy, University of Utah, Salt Lake City, Utah 84112, USA}
\author{Jia-Wei Mei}
\affiliation{Department of Materials Science and Engineering, University of Utah, Salt Lake City, Utah 84112, USA}
\author{ Oleg A. Starykh}
\affiliation{Department of Physics and Astronomy, University of Utah, Salt Lake City, Utah 84112, USA}

\begin{abstract} \label{abstract}
	We describe electron spin resonance in a quantum spin liquid with significant spin-orbit coupling. We find that the resonance directly probes spinon continuum which makes it an efficient and informative probe of exotic excitations of the spin liquid. Specifically, we consider spinon resonance of three different spinon mean-field Hamiltonians, obtained with the help of projective symmetry group analysis, which model a putative quantum spin liquid state of the triangular rare-earth antiferromagnet YbMgGaO$_4$. The band of absorption is found to be very broad and exhibit strong van Hove singularities of single spinon spectrum as well as pronounced polarization dependence.
\end{abstract}
\date{\today}

\maketitle
\section{Introduction}\label{sec:intro}

Electron spin resonance (ESR) and its variants in magnetically ordered systems - ferromagnetic and antiferromagnetic resonances - 
represent one of the most precise and frequently used spectroscopic probes of excitations of magnetic media. The essence of the magnetic resonance technique consists in measuring
absorption of electromagnetic radiation (usually in the microwave range of frequencies) by a sample material which is (typically) subjected to an external static magnetic field.
The absorption is caused by coupling of magnetic degrees of freedom to the magnetic field of the electromagnetic wave. Given very large wavelength of the microwave, the ESR
absorption is driven by zero wavevector (${\bf q} = 0$, or vertical) transitions between states with different $S^z$ projections of magnetic dipole moment on the direction perpendicular
to the magnetic field of the EM radiation.

In a spin system with isotropic exchange, the absorption spectrum of an AC magnetic field is a $\delta$-function peak at the frequency equal to that of the Zeeman energy, 
independently of the exchange interaction strength. This is a consequence of the fact that at ${\bf q} = 0$ EM radiation couples to the total magnetic moment,
which for an $SU(2)$ invariant system commutes with the Hamiltonian \cite{Oshikawa2002}. 
Therefore, any deviation of the absorption spectrum from the $\delta$-function shape implies violation of the spin-rotation symmetry, 
caused either by anisotropic terms in the Hamiltonian (explicit symmetry breaking) or by the development of long-range magnetic order below the critical temperature
(spontaneous symmetry breaking). This is the key reason for ESR's utility.

The goal of our work is to explore applications of ESR to highly entangled phase of magnetic matter - the quantum spin liquid (QSL) \cite{Savary2017}. This intriguing novel quantum state 
manifests itself via non-local elementary excitations - spinons - which behave as fractions of ordinary spin waves. Local spin operator becomes a composite of two or more spinons, which immediately
implies that dynamic spin susceptibility measures multi-spinon continuum. In principle, the best probe of the spinon continuum is provided by inelastic neutron scattering which probes
spinons at finite wave vector ${\bf q}$ and frequency $\omega$. By now several textbook-quality experiments have provided us with unambiguous signatures of multi-particle continua \cite{Dender1997,Coldea2003,Lake2005}.
In practice, however, such state of the art measurements require large high-quality single crystals which quite frequently are not available. 

We posit here that ESR, with its exceptionally high energy resolution, represents an appealing complimentary spectroscopic probe of spinons - {\em spinon magnetic resonance} (SMR). 
The key requirement for turning it into a full-fledged 
probe of spinon dynamics consists in the {\em absence of spin-rotational invariance}. This requirement stems from the mentioned above `insensitivity' of ESR to the details of excitations spectra
in $SU(2)$ invariant magnetic materials. Note that the $SU(2)$ invariance is, at best, a theoretical approximation to the real world materials which {\em always} suffer from some kind of magnetic anisotropy.

Moreover, over the past fifteen years the field of QSL has evolved dramatically {\em away} from the spin-rotational invariance requirement explicit 
in many foundational papers \cite{ANDERSON1987,Affleck1988,Read1991}.
The absence of spin-rotational invariance has evolved from the `real world' annoyance to the virtue \cite{Witczak-Krempa2014,Rau2016,Savary2017}.
Indeed, the first and still the most direct and unambiguous demonstration of the gapless QSL phase came from Kitaev's exact solution of the fully anisotropic honeycomb lattice model \cite{Kitaev2006}
which does not conserve total spin. 

Importantly, a large number of very interesting and not yet understood materials, such as $\alpha$-RuCl$_3$ \cite{Banerjee2016}, YbMgGaO$_4$ \cite{Li2015,Li2015a}, Yb$_2$Ti$_2$O$_7$ \cite{Ross2011,Pan2014}
and many other pyrochlores \cite{Gingras2014}, and even organic BEDT-TTF and BEDT-TSF salts\cite{Winter2017},
showing promising QSL-like features are known to possess significant spin-orbit interaction and are described by spin Hamiltonians with significant asymmetric exchange and pseudo-dipolar terms.
It is precisely this class of low-symmetry spin models we focus on in the present study.

We illustrate our idea by considering spin-liquid state proposed to describe a spin-orbit-coupled triangular lattice Mott insulator YbMgGaO$_4$. The appropriate spin Hamiltonian
has been argued to be that of XXZ model with interactions between nearest (with $J \sim 1$K) and next-nearest neighbors on the triangular lattice 
together with a pseudo-dipolar term \cite{Li2016a,Li2016b,Zhu2017}, 
of $J_{\pm\pm}$ kind in notations of \cite{Ross2011}, between nearest neighbors ($J_{\pm\pm} \sim 0.2$K). 
Most recently, polarized neutron scattering data were interpreted in favor of significant $J_{z\pm}$ interaction \cite{Toth2017}. This Hamiltonian does not conserve total spin ${\bf S}_{\rm tot}$.

Inelastic neutron scattering experiments reveal broad spin excitations continuum \cite{Shen2016,Paddison2017}, consistent with fractionalized QSL with spinon Fermi surface.
At the same time experimental evidence of significant disorder effects \cite{Paddison2017,Li2017,Zhu2017,Xu2016}, capable of masking `pristine' physics of the material, is mounting. 

Our goal here is to add to the ongoing discussion on the nature of the ground state of YbMgGaO$_4$ by pointing out that ESR can serve as a very useful probe of QSL with significant built-in
spin-orbit interactions. We therefore accept spin-liquid hypothesis and focus on fermionic $U(1)$ symmetric spin-liquid ground states, proposed for this material previously \cite{Shen2016, Li2017a}. 
We rely on the well-established projective symmetry group (PSG) analysis of possible $U(1)$ spin liquids \cite{Wen2002,Wen2004,Reuther2014,Li2017a,Bieri1512}.
The spin-orbital nature of the effective spin-1/2 local moment of Yb$^{3+}$ ion 
implies that under the space group symmetry operations both the direction and the position of the local spin are transformed. The symmetry operations include translations 
$T_{1,2}$ along the major axis ${\bf a}_{1,2}$ of the crystal lattice, a rotation $C_2$ by $\pi$ around the in-plane vector ${\bf a}_1 + {\bf a}_2$, 
a counterclockwise rotation $C_3$ by $2\pi/3$ around the lattice site, and the (three-dimensional) inversion $I$ about the lattice site. Following \cite{Li2017a}, it is convenient to combine 
$C_3$ and $I$ operations into a composite one ${\bar C}_6 \equiv C_3^{-1} I$. (Note that the original $C_6$ lattice rotation by $2\pi/6$ around the lattice site is {\em not} the symmetry of
YbMgGaO$_4$ due to alternating - above and below the plane - location of oxygens at the centers of consecutive elementary triangles \cite{Li2015}.)

These symmetries strongly constrain possible $U(1)$ mean-field spinon Hamiltonians and result in 8 different PSG states, of U1A and U1B kind. U1A states maintain periodicity of the original lattice
and their band structure consists of just two spinon bands. 
U1B states are $\pi$-flux states with doubled unit cell. Equivalently, their band structure contains 4 spinon bands. For the sake of simplicity, 
we focus on the U1A family in the following (description of U1B increases algebraic complexity without adding any new essential physics).
The $U(1)$ mean-field spinon Hamiltonian is parameterized by several hopping amplitudes - $t_{1,2}$ describe spin-conserving hopping between the nearest and the next-nearest neighbors
and $t_{1,2}'$ describes analogous non-spin-conserving hops. PSG analysis fixes relative phases between hopping amplitudes on the bonds related by the space group operations
(see Supplement \cite{suppl} for details of the derivation).
The magnitudes of these hoppings are not determined by PSG. This requires a separate variational calculation of the ground state energy which is not attempted here.
We do expect, on physical grounds, that for the spin model with predominant isotropic nearest-neighbor spin exchange and subleading asymmetric $J_{\pm\pm}$ terms, the following
estimate should hold $t_1 > t_1' > t_2 > t_2'$. 

There are four mean-field Hamiltonians in U1A family, labeled by U1A$n_{C_2}n_{\bar{C}_6}$ ($n_{C_2},n_{\bar{C}_6}\in\{0,1\}$).They have the simple form
\begin{equation}
H = \sum_{\bf k} (f^\dagger_{{\bf k} \uparrow}, f^\dagger_{{\bf k} \downarrow}) \left(\begin{matrix} \omega_{\bf k} + \epsilon_{\bf k} & \eta_{\bf k} \\ \eta_{\bf k}^* & \omega_{\bf k} -\epsilon_{\bf k}\end{matrix}\right)  
\left(\begin{matrix} f_{{\bf k} \uparrow} \\ f_{{\bf k} \downarrow} \end{matrix}\right),
\label{eq:1}
\end{equation}
where ${\bf k}$-dependent $\omega_{\bf k}, \epsilon_{\bf k}, \eta_{\bf k}$ are listed in \cite{suppl}. Spin-orbit interaction appears via spin-non-conserving hopping $\eta_k$ in \eqref{eq:1}. The U1A00 state is characterized by finite $\omega_{\bf k}$ and zero $\epsilon_{\bf k}$ and $\eta_{\bf k}$,
while U1A01 and U1A11 have $\omega_{\bf k}  = 0$ and finite $\epsilon_{\bf k}$ and $\eta_{\bf k}$. In the calculations below we set $t_1 =1$ 
and $t_1' = 0.8, t_2' = 0.3$ for U1A11,
while $t_1 =0$ for U1A01 and we choose $t_1' = 1, t_2 = 0.8 t_1', t_2' = 0.4 t_1'$ for it.
U1A10 turns out to be non-physical since its Hamiltonian matrix is zero, $t_{1,2} = t_{1,2}'=0$.
`Accidental' nature of U1A00 state is manifested by the absence of any spin-dependent hopping in its Hamiltonian - this state happens to be more symmetric than the spin Hamiltonian it describes
and is characterized by the large Fermi surface \cite{Shen2016}.

We focus on most physically relevant U1A01 and U1A11 states, for which $\omega_{\bf k}=0$.
The resulting fermion bands are easy to find, $E_{\nu = 1,2}({\bf k}) = (-1)^\nu E({\bf k}) = (-1)^\nu \sqrt{\epsilon_{\bf k}^2 + |\eta_{\bf k}|^2}$.  
U1A11 state possess symmetry-protected Dirac nodes at $\Gamma$ and M points of the hexagonal
Brillouin zone, while U1A01 has additional Dirac nodes at K points as well. 

\begin{figure}
	\includegraphics[scale=0.6]{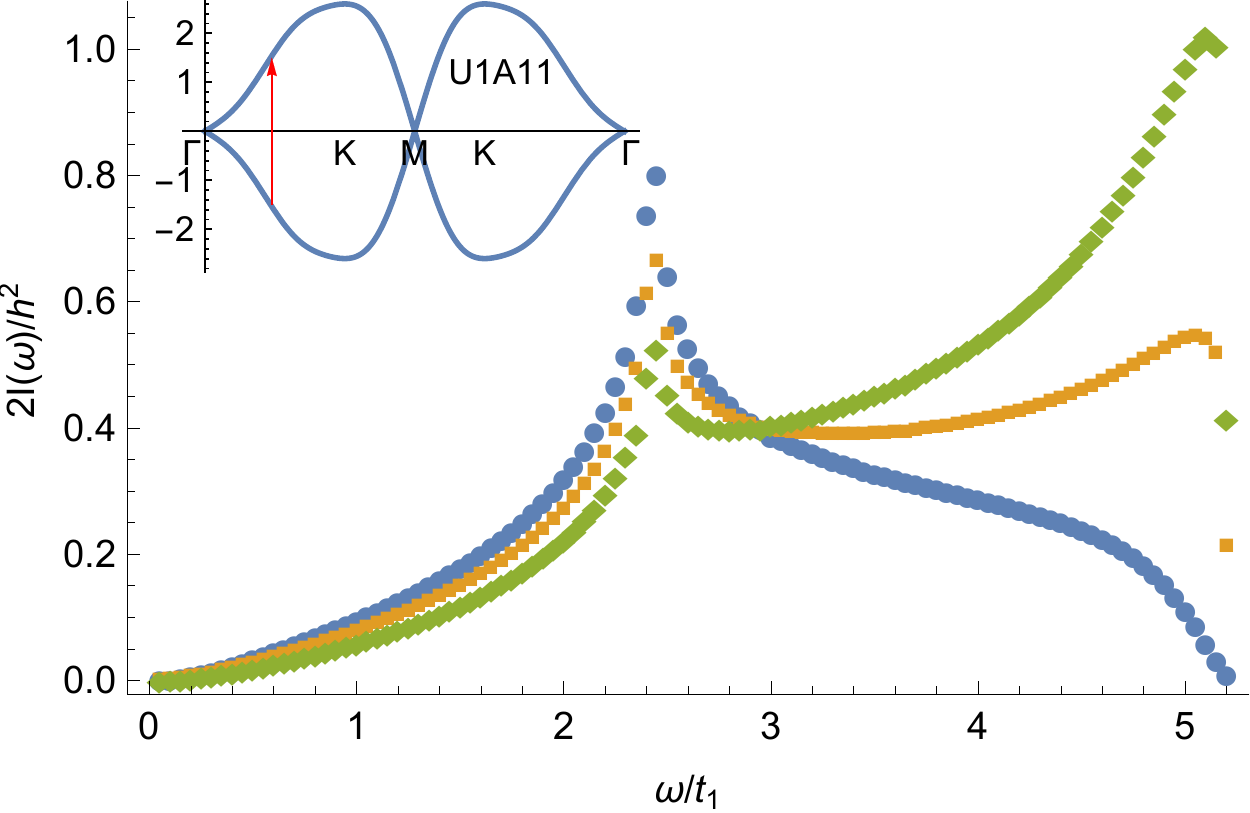}
	\caption{(Color online) Plot of $2 I(\omega)/|h|^2$ vs $\omega/t_1$ for different polarizations $\theta = 0$ (blue dots), $\pi/4$ (orange squares), and $\pi/2$ (green rhombi)  for U1A11 state. The insert shows spinon band structure along the high-symmetry path $\Gamma$-K-M-K-$\Gamma$ in the Brillouin zone.
		Vertical red line illustrates optical transitions between spinon bands.}
	\label{fig:1}
\end{figure}

Interaction with monochromatic radiation linearly polarized along direction $\hat{{\bf n}} = (\sin\theta \cos\phi, \sin\theta \sin\phi, \cos\theta)$ is described by $V(t) = -{\bf h}(t) \cdot {\bf S}_{\rm tot}$, {\em i.e.}
\begin{equation}
V(t) = h e^{-i\omega t} {\bf n} \cdot \frac{1}{2}\sum_{\bf r} (f^\dagger_{{\bf r} \uparrow}, f^\dagger_{{\bf r} \downarrow}) {\bm \sigma} \left(\begin{matrix} f_{{\bf r} \uparrow} \\ f_{{\bf r} \downarrow} \end{matrix}\right)
\label{eq:2}
\end{equation}
Within linear response theory the rate of energy absorption $I(\omega) = -\omega \chi''_{\rm n n}(\omega) |h|^2/2$ is determined by the imaginary part of  ${\bf q}=0$ Fourier transform of the dynamic susceptibility \cite{Oshikawa2002}
$\chi_{\rm n n}(t, {\bf r}) = - i \Theta(t) \langle [{\bf S}_{\bf r}(t) \cdot \hat{{\bf n}}, {\bf S}_{\bf 0}(0) \cdot \hat{{\bf n}}]\rangle$, with $\Theta$ being the Heaviside function. Straightforward calculation gives
\begin{eqnarray}
\chi_{\rm n n}(\omega) &=& \frac{1}{4N}\sum_{\bf k} \frac{n_{{\bf k} \alpha} - n_{{\bf k} \beta}}{\omega + E_\alpha({\bf k}) - E_\beta({\bf k}) + i0} \nonumber\\
&&\times (U_{\bf k}^+ \sigma^a U_{\bf k})_{\alpha \beta} (U_{\bf k}^+ \sigma^b U_{\bf k})_{\beta \alpha} \hat{n}^a \hat{n}^b
\label{eq:3}
\end{eqnarray}
Here $n_{{\bf k} \alpha}$ is the occupation number of the band $\alpha$, $U_{\bf k}$ is unitary diagonalizing matrix connecting spinor of original fermions to that of the band ones,
$f_{{\bf k}, \alpha} = (U_{{\bf k}})_{\alpha\beta} b_{{\bf k},\beta}$, and summation over repeated indices is implied. Eq.\eqref{eq:3} shows that in the spin-degenerate U1A00 state, for which 
$n_{{\bf k} \alpha} = n_{{\bf k} \beta}$, the susceptibility is strictly zero. Therefore, in agreement with general discussion above, no energy absorption occurs in the absence of external magnetic field for this state.
The condition $n_{{\bf k} \alpha} \approx n_{{\bf k} \beta}$ is also satisfied at high temperature of the order of spinon bandwidth (which is of the order of exchange $J$) when spinon resonance
disappears. We therefore expect the width of the resonance to {\em increase} when the temperature is lowered.
It is worth noting that the lowest temperature of ESR study \cite{Li2015a} is $1.8$K, which makes it a high-temperature measurement.

At zero temperature absorption at frequency $\omega$
is possible only via vertical transitions from the filled lower band ($\alpha = 1, n_{{\bf k} 1}=1$) to the empty upper one ($\beta = 2, n_{{\bf k} 2}=0$) and therefore
\begin{eqnarray}
\chi''_{\rm n n}(\omega) &=& -\frac{\pi}{4N}\sum_{\bf k} \delta(\omega - 2 E({\bf k})) (U_{\bf k}^+ \sigma^a U_{\bf k})_{12} \nonumber\\
&& \times  (U_{\bf k}^+ \sigma^b U_{\bf k})_{21} \hat{n}^a \hat{n}^b
\label{eq:4}
\end{eqnarray}
After some algebra, the product of matrix elements $12$ and $21$ of the rotated Pauli matrices in the equation above simplifies to 
\begin{eqnarray}
\chi''_{\rm n n}(\omega) &=& -\frac{\pi}{4N}\sum_{\bf k} \frac{\delta(\omega - 2 E({\bf k}))}{E({\bf k})^2}  \left[(\epsilon_{\bf k}^2 + \eta_{\bf k}''^2) \sin^2\theta\cos^2\phi\right.  \nonumber\\ 
&&\left.+(\epsilon_{\bf k}^2 + \eta_{\bf k}'^2) \sin^2\theta\sin^2\phi  + |\eta_{\bf k}|^2\cos^2\theta \right].
\label{eq:5}
\end{eqnarray}

It can be shown \cite{suppl} that the omitted off-diagonal terms, containing products $\hat{n}^x \hat{n}^y, \hat{n}^x \hat{n}^z$ and $\hat{n}^y \hat{n}^z$, are all zero. Moreover, terms proportional to $\cos^2\phi$ and $\sin^2\phi$
are actually equal, so that the absorption only depends on the azimuthal angle $\theta$ with respect to the normal to the magnetic layer. 

Two features of this result are worth noting. First, the absorption takes place over the wide band of frequencies,
${\rm min}(E) = 0 < \omega/2 < {\rm max}(E)$, which covers the full bandwidth of two-spinon continuum. 
Second, Eq.\eqref{eq:5} describes {\em zero-field absorption}, which does not require any external static magnetic field ${\bf B}$. 
Both of these are direct consequence of the absence of spin conservation in Eq.\eqref{eq:1}. 

{\bf U1A11 state:} Figure~\ref{fig:1} shows scaled absorption intensity, $2 I(\omega)/|h|^2 = - \omega \chi''_{\rm nn}(\omega)$, for different polarizations. 
Polarization dependence is strong.
The plot is obtained by numerical integration 
of \eqref{eq:5}, with frequency steps of $\Delta \omega=0.05$, over the primitive cell of the reciprocal lattice (${\bf k} = (k_1,k_2)$, where $k_{1,2} \in (0,2\pi)$, see \cite{suppl} for details). 
We approximate the delta-function by the Lorenzian $\delta(x) \approx \pi^{-1} d/(d^2 + x^2)$
with $d=0.01$. We checked that $d=0.05$ results in the same outcome.
As expected, and also easy to check analytically,
$\chi''_{\rm nn}(\omega) \sim \omega$ at small frequencies. This is the consequence of Dirac nodes at $\Gamma$ and M points. Behavior near the upper boundary, $\omega \approx 3\sqrt{3}$,
is determined by the vicinity of K point where $\epsilon({\bf K}) = {\text{const}}$ while $\eta({\bf K}) = 0$. As a result, one obtains $\chi''_{\rm zz} \sim 3\sqrt{3} - \omega$,
while at $\theta = \pi/2$ susceptibility terminates discontinuously in a step-like fashion, $\chi''_{\rm nn} = \chi''_{\rm xx} \sim \Theta(3\sqrt{3} - \omega)$. 
The rounding of the step-function behavior in Fig.~\ref{fig:1}, for $\theta \neq 0$, is caused by the finite width of numerical delta-function used in the integration over the Brillouin zone. The peak in the middle of the
absorption band, at $\omega \approx 2.43$, is caused by the van Hove singularity, of the saddle point kind, of $E({\bf k})$ at ${\bf k}_0 = (k_0, 2\pi - k_0)$ and symmetry-related points. Here $k_0 \approx 1.97$
and $E({\bf k}\approx {\bf k}_0) \approx 1.215 + 1.31 (k_1 + k_2)^2 - 0.23 (k_1 - k_2)^2$. The saddle-point produces logarithmically divergent contribution, $\chi''_{\rm nn} \sim \ln|\omega - 2.43|$,
which matches numerical data in Fig.~\ref{fig:1} perfectly.

\begin{figure}
	\includegraphics[scale=0.6]{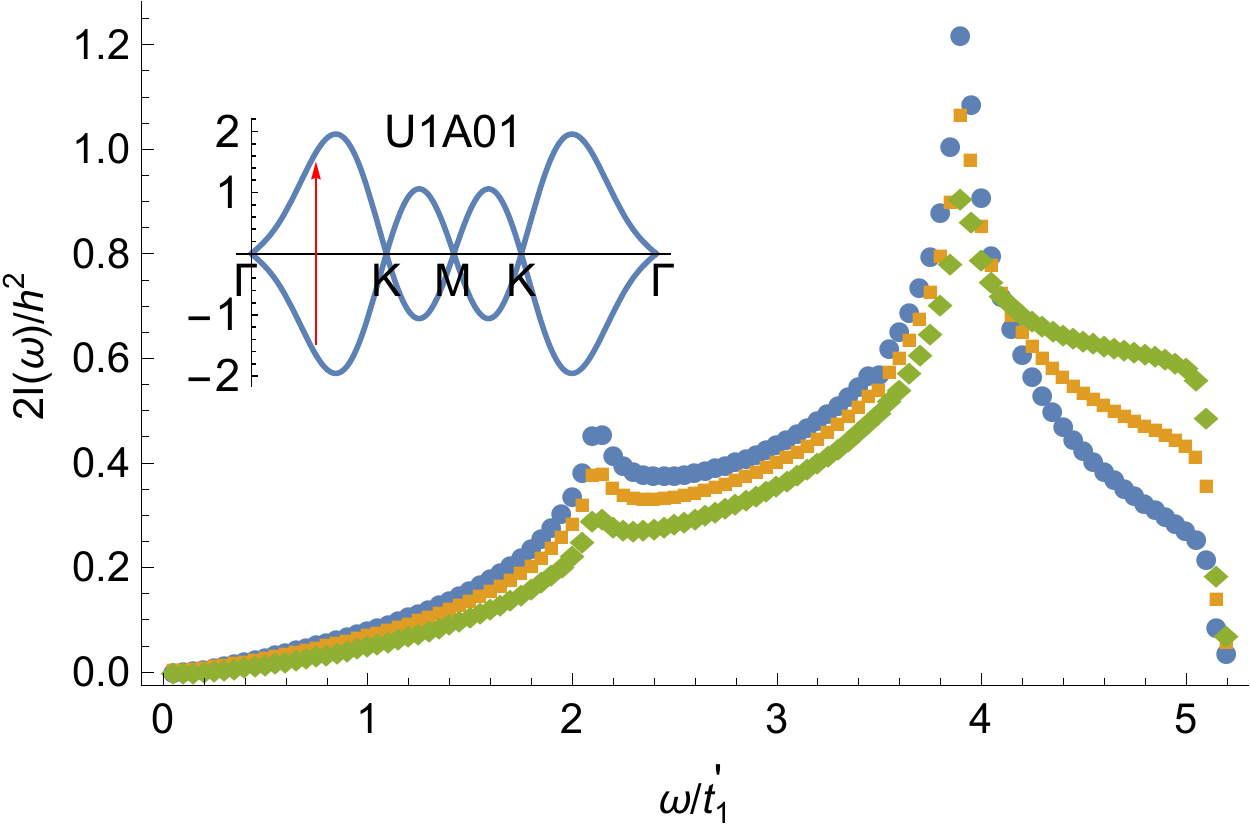}
	\caption{(Color online) Plot of $2 I(\omega)/|h|^2$ vs $\omega/t_1'$ for different polarizations $\theta = 0$ (blue dots), $\pi/4$ (orange squares), and $\pi/2$ (green rhombi) for U1A01 state. The insert shows spinon dispersion along the path in Fig.~\ref{fig:1}.}
	\label{fig:2}
\end{figure}

{\bf U1A01 state:} SMR of this phase is shown in Figure~\ref{fig:2}. It is seen to host two van Hove singularities which can be qualitatively understood as a direct consequence of the
additional, in comparison with U1A11 state, Dirac cone in the spinon dispersion at the K point. The presence of the symmetry-protected node at the K point results in a stronger variation
of spinon dispersion in the Brillouin zone and causes the appearance of additional saddle points.

\begin{figure}
	\includegraphics[scale=0.6]{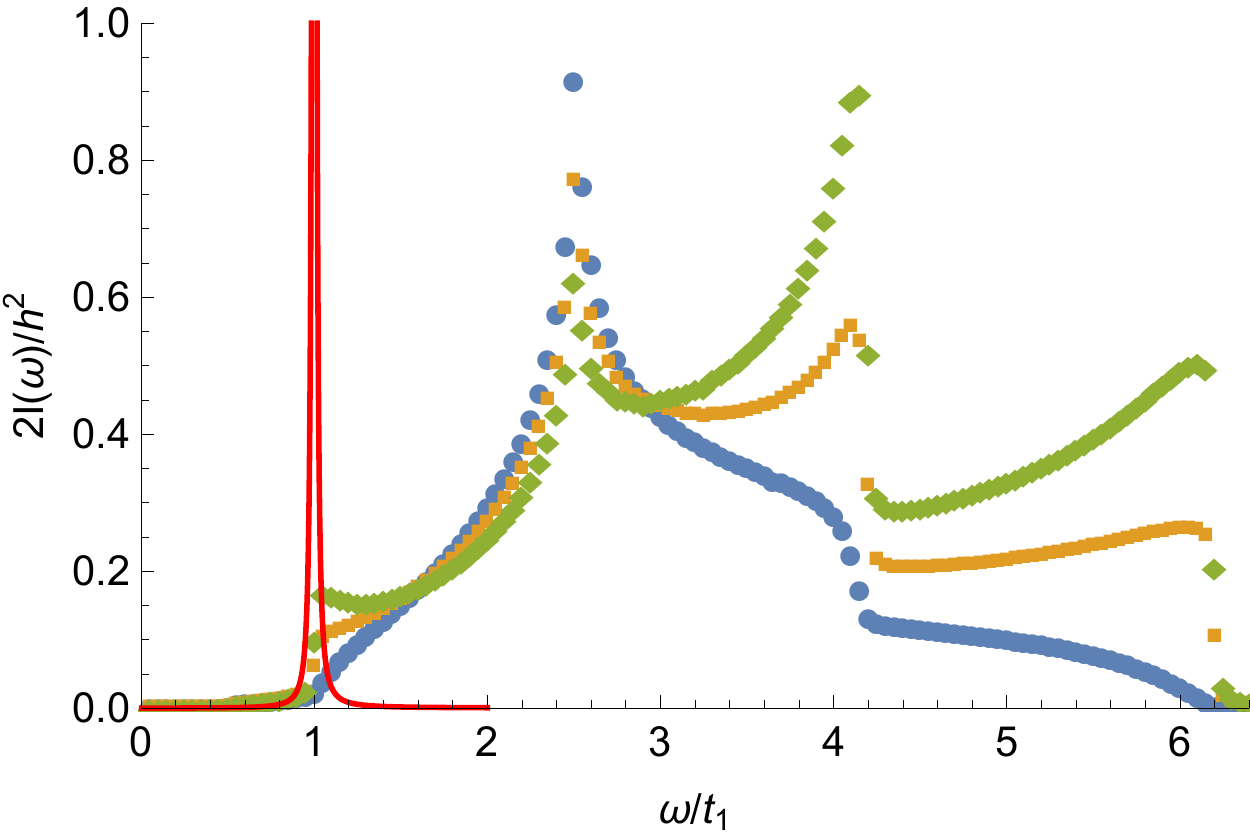}
	\caption{(Color online) Plot of $2 I(\omega)/|h|^2$ vs $\omega/(2t_1)$ for polarizations $\theta = 0$ (blue dots), $\pi/4$ (orange squares), and $\pi/2$ (green rhombi) for U1A11 state in the presence of magnetic field $B_z = 1$ . 
		Note the appearance of strong van Hove singularity at $\omega \approx 4.2 t_1$. Thin red line shows Zeeman response of U1A00 state.}
	\label{fig:3}
\end{figure}

{\bf U1A11 state in magnetic field:} external magnetic field adds further variations to the spinon absorption intensity. We illustrate this with the case
of U1A11 state subject to magnetic field ${\bf B} = B_z \hat{z}$ along the normal to the magnetic layer. It should be noted that PSG analysis underlaying our 
consideration assumes time-reversal (TR) symmetry. Therefore we treat magnetic field perturbatively, by coupling it to the local TR-odd combination
of spinons which is just $B_z S^z_{\bf r} \sim B_z f^\dagger_{{\bf r} \alpha} \sigma^z_{\alpha \beta} f_{{\bf r} \beta}$. Thus magnetic field enters \eqref{eq:1}
via $\epsilon_{\bf k} \to \epsilon_{\bf k} - B_z/2$ and gaps out Dirac nodes. The minimal excitation energy becomes ${\text{min}}(E) = B_z/2$ and
absorption intensity acquires threshold behavior $I(\omega) \sim \Theta(\omega - B_z)$. This behavior is illustrated in Figure \ref{fig:3}, which also shows
development of additional spectral features at $\omega \approx 4.2$, see \cite{suppl}.
In-plane magnetic field lowers symmetry of the spin Hamiltonian further and its consideration is left for future studies.

This unusual response should be contrasted with that of the large-Fermi-surface state U1A00. Here ${\bf B} = B_z \hat{z}$ leads to the Zeeman splitting of spinon up- and down-spin bands $E_\nu = \omega_{\bf k} \mp B_z/2$ and therefore,
according to \eqref{eq:3}, one finds standard result  for magnetically isotropic media $\chi''_{\rm n n}(\omega) \sim \sin^2\theta ~\delta(\omega - B_z)$.
This is consistent with the earlier analysis of \cite{Li1703}, where a weak magnetic field ${\bf B} = B_z \hat{z}$ was added to the mean-field Hamiltonian similarly. Off the $\Gamma$ point, i.e. for ${\bf q}\neq 0$, one finds broad continuum corresponding to the spinon particle-hole excitations \cite{Li1703}.

{\bf Discussion:}
Physical arguments leading to Eq.\eqref{eq:5} are very general and rely on absence of long-range magnetic order, existence of fractionalized elementary excitations, which ensure a continuum-like
response to external probes, and significant  built-in spin-orbit interaction, which leads to non-conservation of spin and makes zero-field absorption possible in a wide range of frequencies. 
All of these are very generic conditions which are satisfied by essentially every model of spin liquids of $U(1)$ and $Z_2$ type (but not by spin-conserving $SU(2)$ ones). 
The restriction to low symmetry spin liquids is not really a handicap as it turned out that the number of possible spin liquids with reduced $U(1)$ and $Z_2$ vastly outnumbers that
of $SU(2)$ symmetric ones \cite{Reuther2014,Dodds2013,Huang2017}. In particular, the
SMR should be present in the celebrated Kitaev's honeycomb model \cite{Kitaev2006}, as was emphasized in dynamic structure calculations of \cite{Knolle2014,Knolle2015,OBrien2016,Smith2016}. 
There too one can see
anisotropic spin structure factor ${\cal S}^{aa}({\bf q}=0,\omega)$, with ${\cal S}^{zz} \neq {\cal S}^{xx/yy}$, and sharp van Hove singularities in the Majorana fermion density of states. 
The similarity is not accidental - it follows from the linear mapping between Majorana and projective spinon representations \cite{Burnell2011,You2012}.
Unlike the situation described here, in the exactly solvable gapless Abelian region dynamic response 
appears above a finite threshold energy (which is the energy cost of creating $Z_2$ fluxes). However generic spin exchange perturbations turn the response gapless \cite{Song2016},
so that ${\cal S}^{aa}({\bf q}=0,\omega) \sim \omega$ at low energy.
Resonant inelastic x-ray (RIXS), Raman scattering and parametric pumping of the $Z_2$ Kitaev spin liquid results in a gapless and extended in energy continuum too \cite{Halasz2016,Knolle2014a,Halasz2017,Zvyagin2017}.

Our theory can be broadly thought of as an extension of one-dimensional theories of ESR in spin chains with Dzyaloshinskii-Moriya interactions 
\cite{Oshikawa2002,Oshikawa1999,Gangadharaiah2008,Karimi2011}. In one dimension, fractionalized nature of spinons is very well established and theories
based on them describe ESR experiments exceedingly well, both in gapless \cite{Zvyagin2005,Povarov2011,Halg2014} and gapped \cite{Glazkov2015,Ozerov2015} settings.

Another important connection is provided by electric dipole spin resonance (EDSR) which describes absorption of EM radiation in {\em conductors} with pronounced spin-orbit interaction
which mediates coupling of AC electric field to the electron spin \cite{Rashba1965}. 
Here, spin-rotational asymmetry
causes strong absorption which is controlled by the real part of optical conductivity \cite{Farid2006,Abanov2012,Glenn2012,Sun2015,Maiti2016,Pokrovsky2017,Bolens2017}.

Somewhat surprisingly, energy absorption due to coupling of spins to AC electric field is also possible in strong Mott insulators, provided 
they are built of frustrated triangular units, in which virtual charge fluctuations produce spin-dependent electric polarization 
\cite{Bulaevskii2008,Potter2013}. Hints of this physics were recently observed in herbertsmithite and $\alpha$-RuCl$_3$ antiferromagnets
\cite{Pilon2013,Laurita2017,Little2017}. 

Simple calculations of SMR presented here are based on mean-field spinon Hamiltonians derived with the help of PSG formalism. They do not include gauge fluctuations which 
undoubtedly are present in the theory. These fluctuations are certain to affect exponents characterizing sharp features of $\chi''_{\rm n n}(\omega)$, such as for example behavior
near the van Hove singularity and/or near lower/upper edge of the two-spinon continuum. 
(Disorder, in the form of Mg/Ga mixing, leads to distribution of g-factors \cite{Li2017} which also broadens magnetic response.)
In addition, by analogy with critical Heisenberg chain \cite{Caux2006}, we expect four-spinon contributions to the
susceptibility to affect the high-frequency behavior. However, these important effects {\em can not reduce} spinon absorption bandwidth and eliminate other outstanding features of the SMR found here.
It should also be noted that SMR is not specific to fermionic spinons and indeed extension of the theory to bosonic PSG is possible as well \cite{Wang2006,Messio2013}.
We therefore conclude that spinon magnetic resonance represents an
efficient and informative probe of exotic excitations of spin-orbit-coupled quantum spin liquids.

\begin{acknowledgments}
	O.A.S. thanks Leon Balents for extensive and insightful discussions of YbMgGaO$_4$ and PSG formalism, Mike Hermele for informative remarks, Sasha Chernyshev for numerous fruitful conversations,
	Natasha Perkins and Dima Pesin for comments on the manuscript.
	Z.-X. L. thanks Yao-Dong Li for helpful discussions.
	O.A.S. is supported by the National Science Foundation grant NSF DMR-1507054.
\end{acknowledgments}

\textbf{Notes Added.} Manuscript \cite{Iaconis2017}, which appeared after our submission, contains detailed comparison of ground state energies of various U(1) PSG states.

\bibliography{Refs.bib}

\begin{thebibliography}{71}%
\makeatletter
\providecommand \@ifxundefined [1]{%
 \@ifx{#1\undefined}
}%
\providecommand \@ifnum [1]{%
 \ifnum #1\expandafter \@firstoftwo
 \else \expandafter \@secondoftwo
 \fi
}%
\providecommand \@ifx [1]{%
 \ifx #1\expandafter \@firstoftwo
 \else \expandafter \@secondoftwo
 \fi
}%
\providecommand \natexlab [1]{#1}%
\providecommand \enquote  [1]{``#1''}%
\providecommand \bibnamefont  [1]{#1}%
\providecommand \bibfnamefont [1]{#1}%
\providecommand \citenamefont [1]{#1}%
\providecommand \href@noop [0]{\@secondoftwo}%
\providecommand \href [0]{\begingroup \@sanitize@url \@href}%
\providecommand \@href[1]{\@@startlink{#1}\@@href}%
\providecommand \@@href[1]{\endgroup#1\@@endlink}%
\providecommand \@sanitize@url [0]{\catcode `\\12\catcode `\$12\catcode
  `\&12\catcode `\#12\catcode `\^12\catcode `\_12\catcode `\%12\relax}%
\providecommand \@@startlink[1]{}%
\providecommand \@@endlink[0]{}%
\providecommand \url  [0]{\begingroup\@sanitize@url \@url }%
\providecommand \@url [1]{\endgroup\@href {#1}{\urlprefix }}%
\providecommand \urlprefix  [0]{URL }%
\providecommand \Eprint [0]{\href }%
\providecommand \doibase [0]{http://dx.doi.org/}%
\providecommand \selectlanguage [0]{\@gobble}%
\providecommand \bibinfo  [0]{\@secondoftwo}%
\providecommand \bibfield  [0]{\@secondoftwo}%
\providecommand \translation [1]{[#1]}%
\providecommand \BibitemOpen [0]{}%
\providecommand \bibitemStop [0]{}%
\providecommand \bibitemNoStop [0]{.\EOS\space}%
\providecommand \EOS [0]{\spacefactor3000\relax}%
\providecommand \BibitemShut  [1]{\csname bibitem#1\endcsname}%
\let\auto@bib@innerbib\@empty
\bibitem [{\citenamefont {Oshikawa}\ and\ \citenamefont
  {Affleck}(2002)}]{Oshikawa2002}%
  \BibitemOpen
  \bibfield  {author} {\bibinfo {author} {\bibfnamefont {M.}~\bibnamefont
  {Oshikawa}}\ and\ \bibinfo {author} {\bibfnamefont {I.}~\bibnamefont
  {Affleck}},\ }\href {\doibase 10.1103/PhysRevB.65.134410} {\bibfield
  {journal} {\bibinfo  {journal} {Phys. Rev. B}\ }\textbf {\bibinfo {volume}
  {65}},\ \bibinfo {pages} {134410} (\bibinfo {year} {2002})}\BibitemShut
  {NoStop}%
\bibitem [{\citenamefont {Savary}\ and\ \citenamefont
  {Balents}(2017)}]{Savary2017}%
  \BibitemOpen
  \bibfield  {author} {\bibinfo {author} {\bibfnamefont {L.}~\bibnamefont
  {Savary}}\ and\ \bibinfo {author} {\bibfnamefont {L.}~\bibnamefont
  {Balents}},\ }\href {http://stacks.iop.org/0034-4885/80/i=1/a=016502}
  {\bibfield  {journal} {\bibinfo  {journal} {Reports on Progress in Physics}\
  }\textbf {\bibinfo {volume} {80}},\ \bibinfo {pages} {016502} (\bibinfo
  {year} {2017})}\BibitemShut {NoStop}%
\bibitem [{\citenamefont {Dender}\ \emph {et~al.}(1997)\citenamefont {Dender},
  \citenamefont {Hammar}, \citenamefont {Reich}, \citenamefont {Broholm},\ and\
  \citenamefont {Aeppli}}]{Dender1997}%
  \BibitemOpen
  \bibfield  {author} {\bibinfo {author} {\bibfnamefont {D.~C.}\ \bibnamefont
  {Dender}}, \bibinfo {author} {\bibfnamefont {P.~R.}\ \bibnamefont {Hammar}},
  \bibinfo {author} {\bibfnamefont {D.~H.}\ \bibnamefont {Reich}}, \bibinfo
  {author} {\bibfnamefont {C.}~\bibnamefont {Broholm}}, \ and\ \bibinfo
  {author} {\bibfnamefont {G.}~\bibnamefont {Aeppli}},\ }\href {\doibase
  10.1103/PhysRevLett.79.1750} {\bibfield  {journal} {\bibinfo  {journal}
  {Phys. Rev. Lett.}\ }\textbf {\bibinfo {volume} {79}},\ \bibinfo {pages}
  {1750} (\bibinfo {year} {1997})}\BibitemShut {NoStop}%
\bibitem [{\citenamefont {Coldea}\ \emph {et~al.}(2003)\citenamefont {Coldea},
  \citenamefont {Tennant},\ and\ \citenamefont {Tylczynski}}]{Coldea2003}%
  \BibitemOpen
  \bibfield  {author} {\bibinfo {author} {\bibfnamefont {R.}~\bibnamefont
  {Coldea}}, \bibinfo {author} {\bibfnamefont {D.~A.}\ \bibnamefont {Tennant}},
  \ and\ \bibinfo {author} {\bibfnamefont {Z.}~\bibnamefont {Tylczynski}},\
  }\href {\doibase 10.1103/PhysRevB.68.134424} {\bibfield  {journal} {\bibinfo
  {journal} {Phys. Rev. B}\ }\textbf {\bibinfo {volume} {68}},\ \bibinfo
  {pages} {134424} (\bibinfo {year} {2003})}\BibitemShut {NoStop}%
\bibitem [{\citenamefont {Lake}\ \emph {et~al.}(2005)\citenamefont {Lake},
  \citenamefont {Tennant}, \citenamefont {Frost},\ and\ \citenamefont
  {Nagler}}]{Lake2005}%
  \BibitemOpen
  \bibfield  {author} {\bibinfo {author} {\bibfnamefont {B.}~\bibnamefont
  {Lake}}, \bibinfo {author} {\bibfnamefont {D.~A.}\ \bibnamefont {Tennant}},
  \bibinfo {author} {\bibfnamefont {C.~D.}\ \bibnamefont {Frost}}, \ and\
  \bibinfo {author} {\bibfnamefont {S.~E.}\ \bibnamefont {Nagler}},\ }\href
  {\doibase 10.1038/nmat1327} {\bibfield  {journal} {\bibinfo  {journal} {Nat
  Mater}\ }\textbf {\bibinfo {volume} {4}},\ \bibinfo {pages} {329} (\bibinfo
  {year} {2005})}\BibitemShut {NoStop}%
\bibitem [{\citenamefont {Anderson}(1987)}]{ANDERSON1987}%
  \BibitemOpen
  \bibfield  {author} {\bibinfo {author} {\bibfnamefont {P.~W.}\ \bibnamefont
  {Anderson}},\ }\href {\doibase 10.1126/science.235.4793.1196} {\bibfield
  {journal} {\bibinfo  {journal} {Science}\ }\textbf {\bibinfo {volume}
  {235}},\ \bibinfo {pages} {1196} (\bibinfo {year} {1987})}\BibitemShut
  {NoStop}%
\bibitem [{\citenamefont {Affleck}\ and\ \citenamefont
  {Marston}(1988)}]{Affleck1988}%
  \BibitemOpen
  \bibfield  {author} {\bibinfo {author} {\bibfnamefont {I.}~\bibnamefont
  {Affleck}}\ and\ \bibinfo {author} {\bibfnamefont {J.~B.}\ \bibnamefont
  {Marston}},\ }\href {\doibase 10.1103/PhysRevB.37.3774} {\bibfield  {journal}
  {\bibinfo  {journal} {Phys. Rev. B}\ }\textbf {\bibinfo {volume} {37}},\
  \bibinfo {pages} {3774} (\bibinfo {year} {1988})}\BibitemShut {NoStop}%
\bibitem [{\citenamefont {Read}\ and\ \citenamefont
  {Sachdev}(1991)}]{Read1991}%
  \BibitemOpen
  \bibfield  {author} {\bibinfo {author} {\bibfnamefont {N.}~\bibnamefont
  {Read}}\ and\ \bibinfo {author} {\bibfnamefont {S.}~\bibnamefont {Sachdev}},\
  }\href {\doibase 10.1103/PhysRevLett.66.1773} {\bibfield  {journal} {\bibinfo
   {journal} {Phys. Rev. Lett.}\ }\textbf {\bibinfo {volume} {66}},\ \bibinfo
  {pages} {1773} (\bibinfo {year} {1991})}\BibitemShut {NoStop}%
\bibitem [{\citenamefont {Witczak-Krempa}\ \emph {et~al.}(2014)\citenamefont
  {Witczak-Krempa}, \citenamefont {Chen}, \citenamefont {Kim},\ and\
  \citenamefont {Balents}}]{Witczak-Krempa2014}%
  \BibitemOpen
  \bibfield  {author} {\bibinfo {author} {\bibfnamefont {W.}~\bibnamefont
  {Witczak-Krempa}}, \bibinfo {author} {\bibfnamefont {G.}~\bibnamefont
  {Chen}}, \bibinfo {author} {\bibfnamefont {Y.~B.}\ \bibnamefont {Kim}}, \
  and\ \bibinfo {author} {\bibfnamefont {L.}~\bibnamefont {Balents}},\ }\href
  {\doibase 10.1146/annurev-conmatphys-020911-125138} {\bibfield  {journal}
  {\bibinfo  {journal} {Annual Review of Condensed Matter Physics}\ }\textbf
  {\bibinfo {volume} {5}},\ \bibinfo {pages} {57} (\bibinfo {year} {2014})},\
  \Eprint
  {http://arxiv.org/abs/http://dx.doi.org/10.1146/annurev-conmatphys-020911-125138}
  {http://dx.doi.org/10.1146/annurev-conmatphys-020911-125138} \BibitemShut
  {NoStop}%
\bibitem [{\citenamefont {Rau}\ \emph {et~al.}(2016)\citenamefont {Rau},
  \citenamefont {Lee},\ and\ \citenamefont {Kee}}]{Rau2016}%
  \BibitemOpen
  \bibfield  {author} {\bibinfo {author} {\bibfnamefont {J.~G.}\ \bibnamefont
  {Rau}}, \bibinfo {author} {\bibfnamefont {E.~K.-H.}\ \bibnamefont {Lee}}, \
  and\ \bibinfo {author} {\bibfnamefont {H.-Y.}\ \bibnamefont {Kee}},\ }\href
  {\doibase 10.1146/annurev-conmatphys-031115-011319} {\bibfield  {journal}
  {\bibinfo  {journal} {Annual Review of Condensed Matter Physics}\ }\textbf
  {\bibinfo {volume} {7}},\ \bibinfo {pages} {195} (\bibinfo {year} {2016})},\
  \Eprint
  {http://arxiv.org/abs/http://dx.doi.org/10.1146/annurev-conmatphys-031115-011319}
  {http://dx.doi.org/10.1146/annurev-conmatphys-031115-011319} \BibitemShut
  {NoStop}%
\bibitem [{\citenamefont {Kitaev}(2006)}]{Kitaev2006}%
  \BibitemOpen
  \bibfield  {author} {\bibinfo {author} {\bibfnamefont {A.}~\bibnamefont
  {Kitaev}},\ }\href {\doibase https://doi.org/10.1016/j.aop.2005.10.005}
  {\bibfield  {journal} {\bibinfo  {journal} {Annals of Physics}\ }\textbf
  {\bibinfo {volume} {321}},\ \bibinfo {pages} {2 } (\bibinfo {year}
  {2006})}\BibitemShut {NoStop}%
\bibitem [{\citenamefont {Banerjee}\ \emph {et~al.}(2016)\citenamefont
  {Banerjee}, \citenamefont {Bridges}, \citenamefont {Yan}, \citenamefont
  {Aczel}, \citenamefont {Li}, \citenamefont {Stone}, \citenamefont {Granroth},
  \citenamefont {Lumsden}, \citenamefont {Yiu}, \citenamefont {Knolle},
  \citenamefont {Bhattacharjee}, \citenamefont {Kovrizhin}, \citenamefont
  {Moessner}, \citenamefont {Tennant}, \citenamefont {Mandrus},\ and\
  \citenamefont {Nagler}}]{Banerjee2016}%
  \BibitemOpen
  \bibfield  {author} {\bibinfo {author} {\bibfnamefont {A.}~\bibnamefont
  {Banerjee}}, \bibinfo {author} {\bibfnamefont {C.~A.}\ \bibnamefont
  {Bridges}}, \bibinfo {author} {\bibfnamefont {J.-Q.}\ \bibnamefont {Yan}},
  \bibinfo {author} {\bibfnamefont {A.~A.}\ \bibnamefont {Aczel}}, \bibinfo
  {author} {\bibfnamefont {L.}~\bibnamefont {Li}}, \bibinfo {author}
  {\bibfnamefont {M.~B.}\ \bibnamefont {Stone}}, \bibinfo {author}
  {\bibfnamefont {G.~E.}\ \bibnamefont {Granroth}}, \bibinfo {author}
  {\bibfnamefont {M.~D.}\ \bibnamefont {Lumsden}}, \bibinfo {author}
  {\bibfnamefont {Y.}~\bibnamefont {Yiu}}, \bibinfo {author} {\bibfnamefont
  {J.}~\bibnamefont {Knolle}}, \bibinfo {author} {\bibfnamefont
  {S.}~\bibnamefont {Bhattacharjee}}, \bibinfo {author} {\bibfnamefont {D.~L.}\
  \bibnamefont {Kovrizhin}}, \bibinfo {author} {\bibfnamefont {R.}~\bibnamefont
  {Moessner}}, \bibinfo {author} {\bibfnamefont {D.~A.}\ \bibnamefont
  {Tennant}}, \bibinfo {author} {\bibfnamefont {D.~G.}\ \bibnamefont
  {Mandrus}}, \ and\ \bibinfo {author} {\bibfnamefont {S.~E.}\ \bibnamefont
  {Nagler}},\ }\href {http://dx.doi.org/10.1038/nmat4604} {\bibfield  {journal}
  {\bibinfo  {journal} {Nat Mater}\ }\textbf {\bibinfo {volume} {15}},\
  \bibinfo {pages} {733} (\bibinfo {year} {2016})}\BibitemShut {NoStop}%
\bibitem [{\citenamefont {Li}\ \emph {et~al.}(2015{\natexlab{a}})\citenamefont
  {Li}, \citenamefont {Liao}, \citenamefont {Zhang}, \citenamefont {Li},
  \citenamefont {Jin}, \citenamefont {Ling}, \citenamefont {Zhang},
  \citenamefont {Zou}, \citenamefont {Pi}, \citenamefont {Yang}, \citenamefont
  {Wang}, \citenamefont {Wu},\ and\ \citenamefont {Zhang}}]{Li2015}%
  \BibitemOpen
  \bibfield  {author} {\bibinfo {author} {\bibfnamefont {Y.}~\bibnamefont
  {Li}}, \bibinfo {author} {\bibfnamefont {H.}~\bibnamefont {Liao}}, \bibinfo
  {author} {\bibfnamefont {Z.}~\bibnamefont {Zhang}}, \bibinfo {author}
  {\bibfnamefont {S.}~\bibnamefont {Li}}, \bibinfo {author} {\bibfnamefont
  {F.}~\bibnamefont {Jin}}, \bibinfo {author} {\bibfnamefont {L.}~\bibnamefont
  {Ling}}, \bibinfo {author} {\bibfnamefont {L.}~\bibnamefont {Zhang}},
  \bibinfo {author} {\bibfnamefont {Y.}~\bibnamefont {Zou}}, \bibinfo {author}
  {\bibfnamefont {L.}~\bibnamefont {Pi}}, \bibinfo {author} {\bibfnamefont
  {Z.}~\bibnamefont {Yang}}, \bibinfo {author} {\bibfnamefont {J.}~\bibnamefont
  {Wang}}, \bibinfo {author} {\bibfnamefont {Z.}~\bibnamefont {Wu}}, \ and\
  \bibinfo {author} {\bibfnamefont {Q.}~\bibnamefont {Zhang}},\ }\href
  {http://dx.doi.org/10.1038/srep16419} {\bibfield  {journal} {\bibinfo
  {journal} {Scientific Reports}\ }\textbf {\bibinfo {volume} {5}},\ \bibinfo
  {pages} {16419} (\bibinfo {year} {2015}{\natexlab{a}})}\BibitemShut {NoStop}%
\bibitem [{\citenamefont {Li}\ \emph {et~al.}(2015{\natexlab{b}})\citenamefont
  {Li}, \citenamefont {Chen}, \citenamefont {Tong}, \citenamefont {Pi},
  \citenamefont {Liu}, \citenamefont {Yang}, \citenamefont {Wang},\ and\
  \citenamefont {Zhang}}]{Li2015a}%
  \BibitemOpen
  \bibfield  {author} {\bibinfo {author} {\bibfnamefont {Y.}~\bibnamefont
  {Li}}, \bibinfo {author} {\bibfnamefont {G.}~\bibnamefont {Chen}}, \bibinfo
  {author} {\bibfnamefont {W.}~\bibnamefont {Tong}}, \bibinfo {author}
  {\bibfnamefont {L.}~\bibnamefont {Pi}}, \bibinfo {author} {\bibfnamefont
  {J.}~\bibnamefont {Liu}}, \bibinfo {author} {\bibfnamefont {Z.}~\bibnamefont
  {Yang}}, \bibinfo {author} {\bibfnamefont {X.}~\bibnamefont {Wang}}, \ and\
  \bibinfo {author} {\bibfnamefont {Q.}~\bibnamefont {Zhang}},\ }\href
  {\doibase 10.1103/PhysRevLett.115.167203} {\bibfield  {journal} {\bibinfo
  {journal} {Phys. Rev. Lett.}\ }\textbf {\bibinfo {volume} {115}},\ \bibinfo
  {pages} {167203} (\bibinfo {year} {2015}{\natexlab{b}})}\BibitemShut
  {NoStop}%
\bibitem [{\citenamefont {Ross}\ \emph {et~al.}(2011)\citenamefont {Ross},
  \citenamefont {Savary}, \citenamefont {Gaulin},\ and\ \citenamefont
  {Balents}}]{Ross2011}%
  \BibitemOpen
  \bibfield  {author} {\bibinfo {author} {\bibfnamefont {K.~A.}\ \bibnamefont
  {Ross}}, \bibinfo {author} {\bibfnamefont {L.}~\bibnamefont {Savary}},
  \bibinfo {author} {\bibfnamefont {B.~D.}\ \bibnamefont {Gaulin}}, \ and\
  \bibinfo {author} {\bibfnamefont {L.}~\bibnamefont {Balents}},\ }\href
  {\doibase 10.1103/PhysRevX.1.021002} {\bibfield  {journal} {\bibinfo
  {journal} {Phys. Rev. X}\ }\textbf {\bibinfo {volume} {1}},\ \bibinfo {pages}
  {021002} (\bibinfo {year} {2011})}\BibitemShut {NoStop}%
\bibitem [{\citenamefont {Pan}\ \emph {et~al.}(2014)\citenamefont {Pan},
  \citenamefont {Kim}, \citenamefont {Ghosh}, \citenamefont {Morris},
  \citenamefont {Ross}, \citenamefont {Kermarrec}, \citenamefont {Gaulin},
  \citenamefont {Koohpayeh}, \citenamefont {Tchernyshyov},\ and\ \citenamefont
  {Armitage}}]{Pan2014}%
  \BibitemOpen
  \bibfield  {author} {\bibinfo {author} {\bibfnamefont {L.}~\bibnamefont
  {Pan}}, \bibinfo {author} {\bibfnamefont {S.~K.}\ \bibnamefont {Kim}},
  \bibinfo {author} {\bibfnamefont {A.}~\bibnamefont {Ghosh}}, \bibinfo
  {author} {\bibfnamefont {C.~M.}\ \bibnamefont {Morris}}, \bibinfo {author}
  {\bibfnamefont {K.~A.}\ \bibnamefont {Ross}}, \bibinfo {author}
  {\bibfnamefont {E.}~\bibnamefont {Kermarrec}}, \bibinfo {author}
  {\bibfnamefont {B.~D.}\ \bibnamefont {Gaulin}}, \bibinfo {author}
  {\bibfnamefont {S.~M.}\ \bibnamefont {Koohpayeh}}, \bibinfo {author}
  {\bibfnamefont {O.}~\bibnamefont {Tchernyshyov}}, \ and\ \bibinfo {author}
  {\bibfnamefont {N.~P.}\ \bibnamefont {Armitage}},\ }\href {\doibase
  10.1038/ncomms5970} {\bibfield  {journal} {\bibinfo  {journal} {Nat.
  Commun.}\ }\textbf {\bibinfo {volume} {5}},\ \bibinfo {pages} {4970}
  (\bibinfo {year} {2014})}\BibitemShut {NoStop}%
\bibitem [{\citenamefont {Gingras}\ and\ \citenamefont
  {McClarty}(2014)}]{Gingras2014}%
  \BibitemOpen
  \bibfield  {author} {\bibinfo {author} {\bibfnamefont {M.~J.~P.}\
  \bibnamefont {Gingras}}\ and\ \bibinfo {author} {\bibfnamefont {P.~A.}\
  \bibnamefont {McClarty}},\ }\href
  {http://stacks.iop.org/0034-4885/77/i=5/a=056501} {\bibfield  {journal}
  {\bibinfo  {journal} {Reports on Progress in Physics}\ }\textbf {\bibinfo
  {volume} {77}},\ \bibinfo {pages} {056501} (\bibinfo {year}
  {2014})}\BibitemShut {NoStop}%
\bibitem [{\citenamefont {Winter}\ \emph {et~al.}(2017)\citenamefont {Winter},
  \citenamefont {Riedl},\ and\ \citenamefont {Valent\'{\i}}}]{Winter2017}%
  \BibitemOpen
  \bibfield  {author} {\bibinfo {author} {\bibfnamefont {S.~M.}\ \bibnamefont
  {Winter}}, \bibinfo {author} {\bibfnamefont {K.}~\bibnamefont {Riedl}}, \
  and\ \bibinfo {author} {\bibfnamefont {R.}~\bibnamefont {Valent\'{\i}}},\
  }\href {\doibase 10.1103/PhysRevB.95.060404} {\bibfield  {journal} {\bibinfo
  {journal} {Phys. Rev. B}\ }\textbf {\bibinfo {volume} {95}},\ \bibinfo
  {pages} {060404} (\bibinfo {year} {2017})}\BibitemShut {NoStop}%
\bibitem [{\citenamefont {Li}\ \emph {et~al.}(2016)\citenamefont {Li},
  \citenamefont {Wang},\ and\ \citenamefont {Chen}}]{Li2016a}%
  \BibitemOpen
  \bibfield  {author} {\bibinfo {author} {\bibfnamefont {Y.-D.}\ \bibnamefont
  {Li}}, \bibinfo {author} {\bibfnamefont {X.}~\bibnamefont {Wang}}, \ and\
  \bibinfo {author} {\bibfnamefont {G.}~\bibnamefont {Chen}},\ }\href {\doibase
  10.1103/PhysRevB.94.035107} {\bibfield  {journal} {\bibinfo  {journal} {Phys.
  Rev. B}\ }\textbf {\bibinfo {volume} {94}},\ \bibinfo {pages} {035107}
  (\bibinfo {year} {2016})}\BibitemShut {NoStop}%
\bibitem [{\citenamefont {{Li}}\ \emph {et~al.}(2016)\citenamefont {{Li}},
  \citenamefont {{Shen}}, \citenamefont {{Li}}, \citenamefont {{Zhao}},\ and\
  \citenamefont {{Chen}}}]{Li2016b}%
  \BibitemOpen
  \bibfield  {author} {\bibinfo {author} {\bibfnamefont {Y.-D.}\ \bibnamefont
  {{Li}}}, \bibinfo {author} {\bibfnamefont {Y.}~\bibnamefont {{Shen}}},
  \bibinfo {author} {\bibfnamefont {Y.}~\bibnamefont {{Li}}}, \bibinfo {author}
  {\bibfnamefont {J.}~\bibnamefont {{Zhao}}}, \ and\ \bibinfo {author}
  {\bibfnamefont {G.}~\bibnamefont {{Chen}}},\ }\href@noop {} {\bibfield
  {journal} {\bibinfo  {journal} {ArXiv e-prints}\ } (\bibinfo {year}
  {2016})},\ \Eprint {http://arxiv.org/abs/1608.06445} {arXiv:1608.06445
  [cond-mat.str-el]} \BibitemShut {NoStop}%
\bibitem [{\citenamefont {{Zhu}}\ \emph {et~al.}(2017)\citenamefont {{Zhu}},
  \citenamefont {{Maksimov}}, \citenamefont {{White}},\ and\ \citenamefont
  {{Chernyshev}}}]{Zhu2017}%
  \BibitemOpen
  \bibfield  {author} {\bibinfo {author} {\bibfnamefont {Z.}~\bibnamefont
  {{Zhu}}}, \bibinfo {author} {\bibfnamefont {P.~A.}\ \bibnamefont
  {{Maksimov}}}, \bibinfo {author} {\bibfnamefont {S.~R.}\ \bibnamefont
  {{White}}}, \ and\ \bibinfo {author} {\bibfnamefont {A.~L.}\ \bibnamefont
  {{Chernyshev}}},\ }\href@noop {} {\bibfield  {journal} {\bibinfo  {journal}
  {ArXiv e-prints}\ } (\bibinfo {year} {2017})},\ \Eprint
  {http://arxiv.org/abs/1703.02971} {arXiv:1703.02971 [cond-mat.str-el]}
  \BibitemShut {NoStop}%
\bibitem [{\citenamefont {{T{\'o}th}}\ \emph {et~al.}(2017)\citenamefont
  {{T{\'o}th}}, \citenamefont {{Rolfs}}, \citenamefont {{Wildes}},\ and\
  \citenamefont {{R{\"u}egg}}}]{Toth2017}%
  \BibitemOpen
  \bibfield  {author} {\bibinfo {author} {\bibfnamefont {S.}~\bibnamefont
  {{T{\'o}th}}}, \bibinfo {author} {\bibfnamefont {K.}~\bibnamefont {{Rolfs}}},
  \bibinfo {author} {\bibfnamefont {A.~R.}\ \bibnamefont {{Wildes}}}, \ and\
  \bibinfo {author} {\bibfnamefont {C.}~\bibnamefont {{R{\"u}egg}}},\
  }\href@noop {} {\bibfield  {journal} {\bibinfo  {journal} {ArXiv e-prints}\ }
  (\bibinfo {year} {2017})},\ \Eprint {http://arxiv.org/abs/1705.05699}
  {arXiv:1705.05699 [cond-mat.str-el]} \BibitemShut {NoStop}%
\bibitem [{\citenamefont {Shen}\ \emph {et~al.}(2016)\citenamefont {Shen},
  \citenamefont {Li}, \citenamefont {Wo}, \citenamefont {Li}, \citenamefont
  {Shen}, \citenamefont {Pan}, \citenamefont {Wang}, \citenamefont {Walker},
  \citenamefont {Steffens}, \citenamefont {Boehm}, \citenamefont {Hao},
  \citenamefont {Quintero-Castro}, \citenamefont {Harriger}, \citenamefont
  {Frontzek}, \citenamefont {Hao}, \citenamefont {Meng}, \citenamefont {Zhang},
  \citenamefont {Chen},\ and\ \citenamefont {Zhao}}]{Shen2016}%
  \BibitemOpen
  \bibfield  {author} {\bibinfo {author} {\bibfnamefont {Y.}~\bibnamefont
  {Shen}}, \bibinfo {author} {\bibfnamefont {Y.-D.}\ \bibnamefont {Li}},
  \bibinfo {author} {\bibfnamefont {H.}~\bibnamefont {Wo}}, \bibinfo {author}
  {\bibfnamefont {Y.}~\bibnamefont {Li}}, \bibinfo {author} {\bibfnamefont
  {S.}~\bibnamefont {Shen}}, \bibinfo {author} {\bibfnamefont {B.}~\bibnamefont
  {Pan}}, \bibinfo {author} {\bibfnamefont {Q.}~\bibnamefont {Wang}}, \bibinfo
  {author} {\bibfnamefont {H.~C.}\ \bibnamefont {Walker}}, \bibinfo {author}
  {\bibfnamefont {P.}~\bibnamefont {Steffens}}, \bibinfo {author}
  {\bibfnamefont {M.}~\bibnamefont {Boehm}}, \bibinfo {author} {\bibfnamefont
  {Y.}~\bibnamefont {Hao}}, \bibinfo {author} {\bibfnamefont {D.~L.}\
  \bibnamefont {Quintero-Castro}}, \bibinfo {author} {\bibfnamefont {L.~W.}\
  \bibnamefont {Harriger}}, \bibinfo {author} {\bibfnamefont {M.~D.}\
  \bibnamefont {Frontzek}}, \bibinfo {author} {\bibfnamefont {L.}~\bibnamefont
  {Hao}}, \bibinfo {author} {\bibfnamefont {S.}~\bibnamefont {Meng}}, \bibinfo
  {author} {\bibfnamefont {Q.}~\bibnamefont {Zhang}}, \bibinfo {author}
  {\bibfnamefont {G.}~\bibnamefont {Chen}}, \ and\ \bibinfo {author}
  {\bibfnamefont {J.}~\bibnamefont {Zhao}},\ }\href
  {http://dx.doi.org/10.1038/nature20614} {\bibfield  {journal} {\bibinfo
  {journal} {Nature}\ }\textbf {\bibinfo {volume} {540}},\ \bibinfo {pages}
  {559} (\bibinfo {year} {2016})}\BibitemShut {NoStop}%
\bibitem [{\citenamefont {Paddison}\ \emph {et~al.}(2017)\citenamefont
  {Paddison}, \citenamefont {Daum}, \citenamefont {Dun}, \citenamefont
  {Ehlers}, \citenamefont {Liu}, \citenamefont {Stone}, \citenamefont {Zhou},\
  and\ \citenamefont {Mourigal}}]{Paddison2017}%
  \BibitemOpen
  \bibfield  {author} {\bibinfo {author} {\bibfnamefont {J.~A.~M.}\
  \bibnamefont {Paddison}}, \bibinfo {author} {\bibfnamefont {M.}~\bibnamefont
  {Daum}}, \bibinfo {author} {\bibfnamefont {Z.}~\bibnamefont {Dun}}, \bibinfo
  {author} {\bibfnamefont {G.}~\bibnamefont {Ehlers}}, \bibinfo {author}
  {\bibfnamefont {Y.}~\bibnamefont {Liu}}, \bibinfo {author} {\bibfnamefont
  {M.~B.}\ \bibnamefont {Stone}}, \bibinfo {author} {\bibfnamefont
  {H.}~\bibnamefont {Zhou}}, \ and\ \bibinfo {author} {\bibfnamefont
  {M.}~\bibnamefont {Mourigal}},\ }\href {http://dx.doi.org/10.1038/nphys3971}
  {\bibfield  {journal} {\bibinfo  {journal} {Nat Phys}\ }\textbf {\bibinfo
  {volume} {13}},\ \bibinfo {pages} {117} (\bibinfo {year} {2017})}\BibitemShut
  {NoStop}%
\bibitem [{\citenamefont {Li}\ \emph {et~al.}(2017{\natexlab{a}})\citenamefont
  {Li}, \citenamefont {Adroja}, \citenamefont {Bewley}, \citenamefont
  {Voneshen}, \citenamefont {Tsirlin}, \citenamefont {Gegenwart},\ and\
  \citenamefont {Zhang}}]{Li2017}%
  \BibitemOpen
  \bibfield  {author} {\bibinfo {author} {\bibfnamefont {Y.}~\bibnamefont
  {Li}}, \bibinfo {author} {\bibfnamefont {D.}~\bibnamefont {Adroja}}, \bibinfo
  {author} {\bibfnamefont {R.~I.}\ \bibnamefont {Bewley}}, \bibinfo {author}
  {\bibfnamefont {D.}~\bibnamefont {Voneshen}}, \bibinfo {author}
  {\bibfnamefont {A.~A.}\ \bibnamefont {Tsirlin}}, \bibinfo {author}
  {\bibfnamefont {P.}~\bibnamefont {Gegenwart}}, \ and\ \bibinfo {author}
  {\bibfnamefont {Q.}~\bibnamefont {Zhang}},\ }\href {\doibase
  10.1103/PhysRevLett.118.107202} {\bibfield  {journal} {\bibinfo  {journal}
  {Phys. Rev. Lett.}\ }\textbf {\bibinfo {volume} {118}},\ \bibinfo {pages}
  {107202} (\bibinfo {year} {2017}{\natexlab{a}})}\BibitemShut {NoStop}%
\bibitem [{\citenamefont {Xu}\ \emph {et~al.}(2016)\citenamefont {Xu},
  \citenamefont {Zhang}, \citenamefont {Li}, \citenamefont {Yu}, \citenamefont
  {Hong}, \citenamefont {Zhang},\ and\ \citenamefont {Li}}]{Xu2016}%
  \BibitemOpen
  \bibfield  {author} {\bibinfo {author} {\bibfnamefont {Y.}~\bibnamefont
  {Xu}}, \bibinfo {author} {\bibfnamefont {J.}~\bibnamefont {Zhang}}, \bibinfo
  {author} {\bibfnamefont {Y.~S.}\ \bibnamefont {Li}}, \bibinfo {author}
  {\bibfnamefont {Y.~J.}\ \bibnamefont {Yu}}, \bibinfo {author} {\bibfnamefont
  {X.~C.}\ \bibnamefont {Hong}}, \bibinfo {author} {\bibfnamefont {Q.~M.}\
  \bibnamefont {Zhang}}, \ and\ \bibinfo {author} {\bibfnamefont {S.~Y.}\
  \bibnamefont {Li}},\ }\href {\doibase 10.1103/PhysRevLett.117.267202}
  {\bibfield  {journal} {\bibinfo  {journal} {Phys. Rev. Lett.}\ }\textbf
  {\bibinfo {volume} {117}},\ \bibinfo {pages} {267202} (\bibinfo {year}
  {2016})}\BibitemShut {NoStop}%
\bibitem [{\citenamefont {Li}\ \emph {et~al.}(2017{\natexlab{b}})\citenamefont
  {Li}, \citenamefont {Lu},\ and\ \citenamefont {Chen}}]{Li2017a}%
  \BibitemOpen
  \bibfield  {author} {\bibinfo {author} {\bibfnamefont {Y.-D.}\ \bibnamefont
  {Li}}, \bibinfo {author} {\bibfnamefont {Y.-M.}\ \bibnamefont {Lu}}, \ and\
  \bibinfo {author} {\bibfnamefont {G.}~\bibnamefont {Chen}},\ }\href {\doibase
  10.1103/PhysRevB.96.054445} {\bibfield  {journal} {\bibinfo  {journal} {Phys.
  Rev. B}\ }\textbf {\bibinfo {volume} {96}},\ \bibinfo {pages} {054445}
  (\bibinfo {year} {2017}{\natexlab{b}})}\BibitemShut {NoStop}%
\bibitem [{\citenamefont {Wen}(2002)}]{Wen2002}%
  \BibitemOpen
  \bibfield  {author} {\bibinfo {author} {\bibfnamefont {X.-G.}\ \bibnamefont
  {Wen}},\ }\href {\doibase 10.1103/PhysRevB.65.165113} {\bibfield  {journal}
  {\bibinfo  {journal} {Phys. Rev. B}\ }\textbf {\bibinfo {volume} {65}},\
  \bibinfo {pages} {165113} (\bibinfo {year} {2002})}\BibitemShut {NoStop}%
\bibitem [{\citenamefont {Wen}(2004)}]{Wen2004}%
  \BibitemOpen
  \bibfield  {author} {\bibinfo {author} {\bibfnamefont {X.~G.}\ \bibnamefont
  {Wen}},\ }\href {https://books.google.com/books?id=llnlrfdR4YgC} {\emph
  {\bibinfo {title} {Quantum Field Theory of Many-Body Systems: From the Origin
  of Sound to an Origin of Light and Electrons}}},\ Oxford Graduate Texts\
  (\bibinfo  {publisher} {OUP Oxford},\ \bibinfo {year} {2004})\BibitemShut
  {NoStop}%
\bibitem [{\citenamefont {Reuther}\ \emph {et~al.}(2014)\citenamefont
  {Reuther}, \citenamefont {Lee},\ and\ \citenamefont {Alicea}}]{Reuther2014}%
  \BibitemOpen
  \bibfield  {author} {\bibinfo {author} {\bibfnamefont {J.}~\bibnamefont
  {Reuther}}, \bibinfo {author} {\bibfnamefont {S.-P.}\ \bibnamefont {Lee}}, \
  and\ \bibinfo {author} {\bibfnamefont {J.}~\bibnamefont {Alicea}},\ }\href
  {\doibase 10.1103/PhysRevB.90.174417} {\bibfield  {journal} {\bibinfo
  {journal} {Phys. Rev. B}\ }\textbf {\bibinfo {volume} {90}},\ \bibinfo
  {pages} {174417} (\bibinfo {year} {2014})}\BibitemShut {NoStop}%
\bibitem [{\citenamefont {Bieri}\ \emph {et~al.}(2016)\citenamefont {Bieri},
  \citenamefont {Lhuillier},\ and\ \citenamefont {Messio}}]{Bieri1512}%
  \BibitemOpen
  \bibfield  {author} {\bibinfo {author} {\bibfnamefont {S.}~\bibnamefont
  {Bieri}}, \bibinfo {author} {\bibfnamefont {C.}~\bibnamefont {Lhuillier}}, \
  and\ \bibinfo {author} {\bibfnamefont {L.}~\bibnamefont {Messio}},\
  }\href@noop {} {\bibfield  {journal} {\bibinfo  {journal} {Phys. Rev. B}\
  }\textbf {\bibinfo {volume} {93}},\ \bibinfo {pages} {094437} (\bibinfo
  {year} {2016})}\BibitemShut {NoStop}%
\bibitem [{sup()}]{suppl}%
  \BibitemOpen
  \href@noop {} {}\bibinfo {note} {Supplemental Material reference}\BibitemShut
  {NoStop}%
\bibitem [{\citenamefont {Li}\ and\ \citenamefont {Chen}(2017)}]{Li1703}%
  \BibitemOpen
  \bibfield  {author} {\bibinfo {author} {\bibfnamefont {Y.-D.}\ \bibnamefont
  {Li}}\ and\ \bibinfo {author} {\bibfnamefont {G.}~\bibnamefont {Chen}},\
  }\href {https://arxiv.org/abs/1703.01876} {\bibfield  {journal} {\bibinfo
  {journal} {ArXiv e-prints}\ } (\bibinfo {year} {2017})},\ \Eprint
  {http://arxiv.org/abs/1703.01876} {arXiv:1703.01876 [cond-mat.str-el]}
  \BibitemShut {NoStop}%
\bibitem [{\citenamefont {Dodds}\ \emph {et~al.}(2013)\citenamefont {Dodds},
  \citenamefont {Bhattacharjee},\ and\ \citenamefont {Kim}}]{Dodds2013}%
  \BibitemOpen
  \bibfield  {author} {\bibinfo {author} {\bibfnamefont {T.}~\bibnamefont
  {Dodds}}, \bibinfo {author} {\bibfnamefont {S.}~\bibnamefont
  {Bhattacharjee}}, \ and\ \bibinfo {author} {\bibfnamefont {Y.~B.}\
  \bibnamefont {Kim}},\ }\href {\doibase 10.1103/PhysRevB.88.224413} {\bibfield
   {journal} {\bibinfo  {journal} {Phys. Rev. B}\ }\textbf {\bibinfo {volume}
  {88}},\ \bibinfo {pages} {224413} (\bibinfo {year} {2013})}\BibitemShut
  {NoStop}%
\bibitem [{\citenamefont {Huang}\ \emph {et~al.}(2017)\citenamefont {Huang},
  \citenamefont {Kim},\ and\ \citenamefont {Lu}}]{Huang2017}%
  \BibitemOpen
  \bibfield  {author} {\bibinfo {author} {\bibfnamefont {B.}~\bibnamefont
  {Huang}}, \bibinfo {author} {\bibfnamefont {Y.~B.}\ \bibnamefont {Kim}}, \
  and\ \bibinfo {author} {\bibfnamefont {Y.-M.}\ \bibnamefont {Lu}},\ }\href
  {\doibase 10.1103/PhysRevB.95.054404} {\bibfield  {journal} {\bibinfo
  {journal} {Phys. Rev. B}\ }\textbf {\bibinfo {volume} {95}},\ \bibinfo
  {pages} {054404} (\bibinfo {year} {2017})}\BibitemShut {NoStop}%
\bibitem [{\citenamefont {Knolle}\ \emph
  {et~al.}(2014{\natexlab{a}})\citenamefont {Knolle}, \citenamefont
  {Kovrizhin}, \citenamefont {Chalker},\ and\ \citenamefont
  {Moessner}}]{Knolle2014}%
  \BibitemOpen
  \bibfield  {author} {\bibinfo {author} {\bibfnamefont {J.}~\bibnamefont
  {Knolle}}, \bibinfo {author} {\bibfnamefont {D.~L.}\ \bibnamefont
  {Kovrizhin}}, \bibinfo {author} {\bibfnamefont {J.~T.}\ \bibnamefont
  {Chalker}}, \ and\ \bibinfo {author} {\bibfnamefont {R.}~\bibnamefont
  {Moessner}},\ }\href {\doibase 10.1103/PhysRevLett.112.207203} {\bibfield
  {journal} {\bibinfo  {journal} {Phys. Rev. Lett.}\ }\textbf {\bibinfo
  {volume} {112}},\ \bibinfo {pages} {207203} (\bibinfo {year}
  {2014}{\natexlab{a}})}\BibitemShut {NoStop}%
\bibitem [{\citenamefont {Knolle}\ \emph {et~al.}(2015)\citenamefont {Knolle},
  \citenamefont {Kovrizhin}, \citenamefont {Chalker},\ and\ \citenamefont
  {Moessner}}]{Knolle2015}%
  \BibitemOpen
  \bibfield  {author} {\bibinfo {author} {\bibfnamefont {J.}~\bibnamefont
  {Knolle}}, \bibinfo {author} {\bibfnamefont {D.~L.}\ \bibnamefont
  {Kovrizhin}}, \bibinfo {author} {\bibfnamefont {J.~T.}\ \bibnamefont
  {Chalker}}, \ and\ \bibinfo {author} {\bibfnamefont {R.}~\bibnamefont
  {Moessner}},\ }\href {\doibase 10.1103/PhysRevB.92.115127} {\bibfield
  {journal} {\bibinfo  {journal} {Phys. Rev. B}\ }\textbf {\bibinfo {volume}
  {92}},\ \bibinfo {pages} {115127} (\bibinfo {year} {2015})}\BibitemShut
  {NoStop}%
\bibitem [{\citenamefont {O'Brien}\ \emph {et~al.}(2016)\citenamefont
  {O'Brien}, \citenamefont {Hermanns},\ and\ \citenamefont
  {Trebst}}]{OBrien2016}%
  \BibitemOpen
  \bibfield  {author} {\bibinfo {author} {\bibfnamefont {K.}~\bibnamefont
  {O'Brien}}, \bibinfo {author} {\bibfnamefont {M.}~\bibnamefont {Hermanns}}, \
  and\ \bibinfo {author} {\bibfnamefont {S.}~\bibnamefont {Trebst}},\ }\href
  {\doibase 10.1103/PhysRevB.93.085101} {\bibfield  {journal} {\bibinfo
  {journal} {Phys. Rev. B}\ }\textbf {\bibinfo {volume} {93}},\ \bibinfo
  {pages} {085101} (\bibinfo {year} {2016})}\BibitemShut {NoStop}%
\bibitem [{\citenamefont {Smith}\ \emph {et~al.}(2016)\citenamefont {Smith},
  \citenamefont {Knolle}, \citenamefont {Kovrizhin}, \citenamefont {Chalker},\
  and\ \citenamefont {Moessner}}]{Smith2016}%
  \BibitemOpen
  \bibfield  {author} {\bibinfo {author} {\bibfnamefont {A.}~\bibnamefont
  {Smith}}, \bibinfo {author} {\bibfnamefont {J.}~\bibnamefont {Knolle}},
  \bibinfo {author} {\bibfnamefont {D.~L.}\ \bibnamefont {Kovrizhin}}, \bibinfo
  {author} {\bibfnamefont {J.~T.}\ \bibnamefont {Chalker}}, \ and\ \bibinfo
  {author} {\bibfnamefont {R.}~\bibnamefont {Moessner}},\ }\href {\doibase
  10.1103/PhysRevB.93.235146} {\bibfield  {journal} {\bibinfo  {journal} {Phys.
  Rev. B}\ }\textbf {\bibinfo {volume} {93}},\ \bibinfo {pages} {235146}
  (\bibinfo {year} {2016})}\BibitemShut {NoStop}%
\bibitem [{\citenamefont {Burnell}\ and\ \citenamefont
  {Nayak}(2011)}]{Burnell2011}%
  \BibitemOpen
  \bibfield  {author} {\bibinfo {author} {\bibfnamefont {F.~J.}\ \bibnamefont
  {Burnell}}\ and\ \bibinfo {author} {\bibfnamefont {C.}~\bibnamefont
  {Nayak}},\ }\href {\doibase 10.1103/PhysRevB.84.125125} {\bibfield  {journal}
  {\bibinfo  {journal} {Phys. Rev. B}\ }\textbf {\bibinfo {volume} {84}},\
  \bibinfo {pages} {125125} (\bibinfo {year} {2011})}\BibitemShut {NoStop}%
\bibitem [{\citenamefont {You}\ \emph {et~al.}(2012)\citenamefont {You},
  \citenamefont {Kimchi},\ and\ \citenamefont {Vishwanath}}]{You2012}%
  \BibitemOpen
  \bibfield  {author} {\bibinfo {author} {\bibfnamefont {Y.-Z.}\ \bibnamefont
  {You}}, \bibinfo {author} {\bibfnamefont {I.}~\bibnamefont {Kimchi}}, \ and\
  \bibinfo {author} {\bibfnamefont {A.}~\bibnamefont {Vishwanath}},\ }\href
  {\doibase 10.1103/PhysRevB.86.085145} {\bibfield  {journal} {\bibinfo
  {journal} {Phys. Rev. B}\ }\textbf {\bibinfo {volume} {86}},\ \bibinfo
  {pages} {085145} (\bibinfo {year} {2012})}\BibitemShut {NoStop}%
\bibitem [{\citenamefont {Song}\ \emph {et~al.}(2016)\citenamefont {Song},
  \citenamefont {You},\ and\ \citenamefont {Balents}}]{Song2016}%
  \BibitemOpen
  \bibfield  {author} {\bibinfo {author} {\bibfnamefont {X.-Y.}\ \bibnamefont
  {Song}}, \bibinfo {author} {\bibfnamefont {Y.-Z.}\ \bibnamefont {You}}, \
  and\ \bibinfo {author} {\bibfnamefont {L.}~\bibnamefont {Balents}},\ }\href
  {\doibase 10.1103/PhysRevLett.117.037209} {\bibfield  {journal} {\bibinfo
  {journal} {Phys. Rev. Lett.}\ }\textbf {\bibinfo {volume} {117}},\ \bibinfo
  {pages} {037209} (\bibinfo {year} {2016})}\BibitemShut {NoStop}%
\bibitem [{\citenamefont {Hal\'asz}\ \emph {et~al.}(2016)\citenamefont
  {Hal\'asz}, \citenamefont {Perkins},\ and\ \citenamefont {van~den
  Brink}}]{Halasz2016}%
  \BibitemOpen
  \bibfield  {author} {\bibinfo {author} {\bibfnamefont {G.~B.}\ \bibnamefont
  {Hal\'asz}}, \bibinfo {author} {\bibfnamefont {N.~B.}\ \bibnamefont
  {Perkins}}, \ and\ \bibinfo {author} {\bibfnamefont {J.}~\bibnamefont
  {van~den Brink}},\ }\href {\doibase 10.1103/PhysRevLett.117.127203}
  {\bibfield  {journal} {\bibinfo  {journal} {Phys. Rev. Lett.}\ }\textbf
  {\bibinfo {volume} {117}},\ \bibinfo {pages} {127203} (\bibinfo {year}
  {2016})}\BibitemShut {NoStop}%
\bibitem [{\citenamefont {Knolle}\ \emph
  {et~al.}(2014{\natexlab{b}})\citenamefont {Knolle}, \citenamefont {Chern},
  \citenamefont {Kovrizhin}, \citenamefont {Moessner},\ and\ \citenamefont
  {Perkins}}]{Knolle2014a}%
  \BibitemOpen
  \bibfield  {author} {\bibinfo {author} {\bibfnamefont {J.}~\bibnamefont
  {Knolle}}, \bibinfo {author} {\bibfnamefont {G.-W.}\ \bibnamefont {Chern}},
  \bibinfo {author} {\bibfnamefont {D.~L.}\ \bibnamefont {Kovrizhin}}, \bibinfo
  {author} {\bibfnamefont {R.}~\bibnamefont {Moessner}}, \ and\ \bibinfo
  {author} {\bibfnamefont {N.~B.}\ \bibnamefont {Perkins}},\ }\href {\doibase
  10.1103/PhysRevLett.113.187201} {\bibfield  {journal} {\bibinfo  {journal}
  {Phys. Rev. Lett.}\ }\textbf {\bibinfo {volume} {113}},\ \bibinfo {pages}
  {187201} (\bibinfo {year} {2014}{\natexlab{b}})}\BibitemShut {NoStop}%
\bibitem [{\citenamefont {{Hal{\'a}sz}}\ \emph {et~al.}(2017)\citenamefont
  {{Hal{\'a}sz}}, \citenamefont {{Perreault}},\ and\ \citenamefont
  {{Perkins}}}]{Halasz2017}%
  \BibitemOpen
  \bibfield  {author} {\bibinfo {author} {\bibfnamefont {G.~B.}\ \bibnamefont
  {{Hal{\'a}sz}}}, \bibinfo {author} {\bibfnamefont {B.}~\bibnamefont
  {{Perreault}}}, \ and\ \bibinfo {author} {\bibfnamefont {N.~B.}\ \bibnamefont
  {{Perkins}}},\ }\href@noop {} {\bibfield  {journal} {\bibinfo  {journal}
  {ArXiv e-prints}\ } (\bibinfo {year} {2017})},\ \Eprint
  {http://arxiv.org/abs/1705.05894} {arXiv:1705.05894 [cond-mat.str-el]}
  \BibitemShut {NoStop}%
\bibitem [{\citenamefont {Zvyagin}(2017)}]{Zvyagin2017}%
  \BibitemOpen
  \bibfield  {author} {\bibinfo {author} {\bibfnamefont {A.~A.}\ \bibnamefont
  {Zvyagin}},\ }\href {\doibase 10.1103/PhysRevB.95.064428} {\bibfield
  {journal} {\bibinfo  {journal} {Phys. Rev. B}\ }\textbf {\bibinfo {volume}
  {95}},\ \bibinfo {pages} {064428} (\bibinfo {year} {2017})}\BibitemShut
  {NoStop}%
\bibitem [{\citenamefont {Oshikawa}\ and\ \citenamefont
  {Affleck}(1999)}]{Oshikawa1999}%
  \BibitemOpen
  \bibfield  {author} {\bibinfo {author} {\bibfnamefont {M.}~\bibnamefont
  {Oshikawa}}\ and\ \bibinfo {author} {\bibfnamefont {I.}~\bibnamefont
  {Affleck}},\ }\href {\doibase 10.1103/PhysRevLett.82.5136} {\bibfield
  {journal} {\bibinfo  {journal} {Phys. Rev. Lett.}\ }\textbf {\bibinfo
  {volume} {82}},\ \bibinfo {pages} {5136} (\bibinfo {year}
  {1999})}\BibitemShut {NoStop}%
\bibitem [{\citenamefont {Gangadharaiah}\ \emph {et~al.}(2008)\citenamefont
  {Gangadharaiah}, \citenamefont {Sun},\ and\ \citenamefont
  {Starykh}}]{Gangadharaiah2008}%
  \BibitemOpen
  \bibfield  {author} {\bibinfo {author} {\bibfnamefont {S.}~\bibnamefont
  {Gangadharaiah}}, \bibinfo {author} {\bibfnamefont {J.}~\bibnamefont {Sun}},
  \ and\ \bibinfo {author} {\bibfnamefont {O.~A.}\ \bibnamefont {Starykh}},\
  }\href {\doibase 10.1103/PhysRevB.78.054436} {\bibfield  {journal} {\bibinfo
  {journal} {Phys. Rev. B}\ }\textbf {\bibinfo {volume} {78}},\ \bibinfo
  {pages} {054436} (\bibinfo {year} {2008})}\BibitemShut {NoStop}%
\bibitem [{\citenamefont {Karimi}\ and\ \citenamefont
  {Affleck}(2011)}]{Karimi2011}%
  \BibitemOpen
  \bibfield  {author} {\bibinfo {author} {\bibfnamefont {H.}~\bibnamefont
  {Karimi}}\ and\ \bibinfo {author} {\bibfnamefont {I.}~\bibnamefont
  {Affleck}},\ }\href {\doibase 10.1103/PhysRevB.84.174420} {\bibfield
  {journal} {\bibinfo  {journal} {Phys. Rev. B}\ }\textbf {\bibinfo {volume}
  {84}},\ \bibinfo {pages} {174420} (\bibinfo {year} {2011})}\BibitemShut
  {NoStop}%
\bibitem [{\citenamefont {Zvyagin}\ \emph {et~al.}(2005)\citenamefont
  {Zvyagin}, \citenamefont {Kolezhuk}, \citenamefont {Krzystek},\ and\
  \citenamefont {Feyerherm}}]{Zvyagin2005}%
  \BibitemOpen
  \bibfield  {author} {\bibinfo {author} {\bibfnamefont {S.~A.}\ \bibnamefont
  {Zvyagin}}, \bibinfo {author} {\bibfnamefont {A.~K.}\ \bibnamefont
  {Kolezhuk}}, \bibinfo {author} {\bibfnamefont {J.}~\bibnamefont {Krzystek}},
  \ and\ \bibinfo {author} {\bibfnamefont {R.}~\bibnamefont {Feyerherm}},\
  }\href {\doibase 10.1103/PhysRevLett.95.017207} {\bibfield  {journal}
  {\bibinfo  {journal} {Phys. Rev. Lett.}\ }\textbf {\bibinfo {volume} {95}},\
  \bibinfo {pages} {017207} (\bibinfo {year} {2005})}\BibitemShut {NoStop}%
\bibitem [{\citenamefont {Povarov}\ \emph {et~al.}(2011)\citenamefont
  {Povarov}, \citenamefont {Smirnov}, \citenamefont {Starykh}, \citenamefont
  {Petrov},\ and\ \citenamefont {Shapiro}}]{Povarov2011}%
  \BibitemOpen
  \bibfield  {author} {\bibinfo {author} {\bibfnamefont {K.~Y.}\ \bibnamefont
  {Povarov}}, \bibinfo {author} {\bibfnamefont {A.~I.}\ \bibnamefont
  {Smirnov}}, \bibinfo {author} {\bibfnamefont {O.~A.}\ \bibnamefont
  {Starykh}}, \bibinfo {author} {\bibfnamefont {S.~V.}\ \bibnamefont {Petrov}},
  \ and\ \bibinfo {author} {\bibfnamefont {A.~Y.}\ \bibnamefont {Shapiro}},\
  }\href {\doibase 10.1103/PhysRevLett.107.037204} {\bibfield  {journal}
  {\bibinfo  {journal} {Phys. Rev. Lett.}\ }\textbf {\bibinfo {volume} {107}},\
  \bibinfo {pages} {037204} (\bibinfo {year} {2011})}\BibitemShut {NoStop}%
\bibitem [{\citenamefont {H\"alg}\ \emph {et~al.}(2014)\citenamefont {H\"alg},
  \citenamefont {Lorenz}, \citenamefont {Povarov}, \citenamefont {M\aa{}nsson},
  \citenamefont {Skourski},\ and\ \citenamefont {Zheludev}}]{Halg2014}%
  \BibitemOpen
  \bibfield  {author} {\bibinfo {author} {\bibfnamefont {M.}~\bibnamefont
  {H\"alg}}, \bibinfo {author} {\bibfnamefont {W.~E.~A.}\ \bibnamefont
  {Lorenz}}, \bibinfo {author} {\bibfnamefont {K.~Y.}\ \bibnamefont {Povarov}},
  \bibinfo {author} {\bibfnamefont {M.}~\bibnamefont {M\aa{}nsson}}, \bibinfo
  {author} {\bibfnamefont {Y.}~\bibnamefont {Skourski}}, \ and\ \bibinfo
  {author} {\bibfnamefont {A.}~\bibnamefont {Zheludev}},\ }\href {\doibase
  10.1103/PhysRevB.90.174413} {\bibfield  {journal} {\bibinfo  {journal} {Phys.
  Rev. B}\ }\textbf {\bibinfo {volume} {90}},\ \bibinfo {pages} {174413}
  (\bibinfo {year} {2014})}\BibitemShut {NoStop}%
\bibitem [{\citenamefont {Glazkov}\ \emph {et~al.}(2015)\citenamefont
  {Glazkov}, \citenamefont {Fayzullin}, \citenamefont {Krasnikova},
  \citenamefont {Skoblin}, \citenamefont {Schmidiger}, \citenamefont
  {M\"uhlbauer},\ and\ \citenamefont {Zheludev}}]{Glazkov2015}%
  \BibitemOpen
  \bibfield  {author} {\bibinfo {author} {\bibfnamefont {V.~N.}\ \bibnamefont
  {Glazkov}}, \bibinfo {author} {\bibfnamefont {M.}~\bibnamefont {Fayzullin}},
  \bibinfo {author} {\bibfnamefont {Y.}~\bibnamefont {Krasnikova}}, \bibinfo
  {author} {\bibfnamefont {G.}~\bibnamefont {Skoblin}}, \bibinfo {author}
  {\bibfnamefont {D.}~\bibnamefont {Schmidiger}}, \bibinfo {author}
  {\bibfnamefont {S.}~\bibnamefont {M\"uhlbauer}}, \ and\ \bibinfo {author}
  {\bibfnamefont {A.}~\bibnamefont {Zheludev}},\ }\href {\doibase
  10.1103/PhysRevB.92.184403} {\bibfield  {journal} {\bibinfo  {journal} {Phys.
  Rev. B}\ }\textbf {\bibinfo {volume} {92}},\ \bibinfo {pages} {184403}
  (\bibinfo {year} {2015})}\BibitemShut {NoStop}%
\bibitem [{\citenamefont {Ozerov}\ \emph {et~al.}(2015)\citenamefont {Ozerov},
  \citenamefont {Maksymenko}, \citenamefont {Wosnitza}, \citenamefont
  {Honecker}, \citenamefont {Landee}, \citenamefont {Turnbull}, \citenamefont
  {Furuya}, \citenamefont {Giamarchi},\ and\ \citenamefont
  {Zvyagin}}]{Ozerov2015}%
  \BibitemOpen
  \bibfield  {author} {\bibinfo {author} {\bibfnamefont {M.}~\bibnamefont
  {Ozerov}}, \bibinfo {author} {\bibfnamefont {M.}~\bibnamefont {Maksymenko}},
  \bibinfo {author} {\bibfnamefont {J.}~\bibnamefont {Wosnitza}}, \bibinfo
  {author} {\bibfnamefont {A.}~\bibnamefont {Honecker}}, \bibinfo {author}
  {\bibfnamefont {C.~P.}\ \bibnamefont {Landee}}, \bibinfo {author}
  {\bibfnamefont {M.~M.}\ \bibnamefont {Turnbull}}, \bibinfo {author}
  {\bibfnamefont {S.~C.}\ \bibnamefont {Furuya}}, \bibinfo {author}
  {\bibfnamefont {T.}~\bibnamefont {Giamarchi}}, \ and\ \bibinfo {author}
  {\bibfnamefont {S.~A.}\ \bibnamefont {Zvyagin}},\ }\href {\doibase
  10.1103/PhysRevB.92.241113} {\bibfield  {journal} {\bibinfo  {journal} {Phys.
  Rev. B}\ }\textbf {\bibinfo {volume} {92}},\ \bibinfo {pages} {241113}
  (\bibinfo {year} {2015})}\BibitemShut {NoStop}%
\bibitem [{\citenamefont {Rashba}(1965)}]{Rashba1965}%
  \BibitemOpen
  \bibfield  {author} {\bibinfo {author} {\bibfnamefont {E.~I.}\ \bibnamefont
  {Rashba}},\ }\href {http://stacks.iop.org/0038-5670/7/i=6/a=R05} {\bibfield
  {journal} {\bibinfo  {journal} {Soviet Physics Uspekhi}\ }\textbf {\bibinfo
  {volume} {7}},\ \bibinfo {pages} {823} (\bibinfo {year} {1965})}\BibitemShut
  {NoStop}%
\bibitem [{\citenamefont {Farid}\ and\ \citenamefont
  {Mishchenko}(2006)}]{Farid2006}%
  \BibitemOpen
  \bibfield  {author} {\bibinfo {author} {\bibfnamefont {A.-K.}\ \bibnamefont
  {Farid}}\ and\ \bibinfo {author} {\bibfnamefont {E.~G.}\ \bibnamefont
  {Mishchenko}},\ }\href {\doibase 10.1103/PhysRevLett.97.096604} {\bibfield
  {journal} {\bibinfo  {journal} {Phys. Rev. Lett.}\ }\textbf {\bibinfo
  {volume} {97}},\ \bibinfo {pages} {096604} (\bibinfo {year}
  {2006})}\BibitemShut {NoStop}%
\bibitem [{\citenamefont {Abanov}\ \emph {et~al.}(2012)\citenamefont {Abanov},
  \citenamefont {Pokrovsky}, \citenamefont {Saslow},\ and\ \citenamefont
  {Zhou}}]{Abanov2012}%
  \BibitemOpen
  \bibfield  {author} {\bibinfo {author} {\bibfnamefont {A.}~\bibnamefont
  {Abanov}}, \bibinfo {author} {\bibfnamefont {V.~L.}\ \bibnamefont
  {Pokrovsky}}, \bibinfo {author} {\bibfnamefont {W.~M.}\ \bibnamefont
  {Saslow}}, \ and\ \bibinfo {author} {\bibfnamefont {P.}~\bibnamefont
  {Zhou}},\ }\href {\doibase 10.1103/PhysRevB.85.085311} {\bibfield  {journal}
  {\bibinfo  {journal} {Phys. Rev. B}\ }\textbf {\bibinfo {volume} {85}},\
  \bibinfo {pages} {085311} (\bibinfo {year} {2012})}\BibitemShut {NoStop}%
\bibitem [{\citenamefont {Glenn}\ \emph {et~al.}(2012)\citenamefont {Glenn},
  \citenamefont {Starykh},\ and\ \citenamefont {Raikh}}]{Glenn2012}%
  \BibitemOpen
  \bibfield  {author} {\bibinfo {author} {\bibfnamefont {R.}~\bibnamefont
  {Glenn}}, \bibinfo {author} {\bibfnamefont {O.~A.}\ \bibnamefont {Starykh}},
  \ and\ \bibinfo {author} {\bibfnamefont {M.~E.}\ \bibnamefont {Raikh}},\
  }\href {\doibase 10.1103/PhysRevB.86.024423} {\bibfield  {journal} {\bibinfo
  {journal} {Phys. Rev. B}\ }\textbf {\bibinfo {volume} {86}},\ \bibinfo
  {pages} {024423} (\bibinfo {year} {2012})}\BibitemShut {NoStop}%
\bibitem [{\citenamefont {Sun}\ and\ \citenamefont
  {Pokrovsky}(2015)}]{Sun2015}%
  \BibitemOpen
  \bibfield  {author} {\bibinfo {author} {\bibfnamefont {C.}~\bibnamefont
  {Sun}}\ and\ \bibinfo {author} {\bibfnamefont {V.~L.}\ \bibnamefont
  {Pokrovsky}},\ }\href {\doibase 10.1103/PhysRevB.91.161305} {\bibfield
  {journal} {\bibinfo  {journal} {Phys. Rev. B}\ }\textbf {\bibinfo {volume}
  {91}},\ \bibinfo {pages} {161305} (\bibinfo {year} {2015})}\BibitemShut
  {NoStop}%
\bibitem [{\citenamefont {Maiti}\ \emph {et~al.}(2016)\citenamefont {Maiti},
  \citenamefont {Imran},\ and\ \citenamefont {Maslov}}]{Maiti2016}%
  \BibitemOpen
  \bibfield  {author} {\bibinfo {author} {\bibfnamefont {S.}~\bibnamefont
  {Maiti}}, \bibinfo {author} {\bibfnamefont {M.}~\bibnamefont {Imran}}, \ and\
  \bibinfo {author} {\bibfnamefont {D.~L.}\ \bibnamefont {Maslov}},\ }\href
  {\doibase 10.1103/PhysRevB.93.045134} {\bibfield  {journal} {\bibinfo
  {journal} {Phys. Rev. B}\ }\textbf {\bibinfo {volume} {93}},\ \bibinfo
  {pages} {045134} (\bibinfo {year} {2016})}\BibitemShut {NoStop}%
\bibitem [{\citenamefont {Pokrovsky}(2017)}]{Pokrovsky2017}%
  \BibitemOpen
  \bibfield  {author} {\bibinfo {author} {\bibfnamefont {V.~L.}\ \bibnamefont
  {Pokrovsky}},\ }\href {\doibase 10.1063/1.4976632} {\bibfield  {journal}
  {\bibinfo  {journal} {Low Temperature Physics}\ }\textbf {\bibinfo {volume}
  {43}},\ \bibinfo {pages} {211} (\bibinfo {year} {2017})},\ \Eprint
  {http://arxiv.org/abs/http://dx.doi.org/10.1063/1.4976632}
  {http://dx.doi.org/10.1063/1.4976632} \BibitemShut {NoStop}%
\bibitem [{\citenamefont {{Bolens}}\ \emph {et~al.}(2017)\citenamefont
  {{Bolens}}, \citenamefont {{Katsura}}, \citenamefont {{Ogata}},\ and\
  \citenamefont {{Miyashita}}}]{Bolens2017}%
  \BibitemOpen
  \bibfield  {author} {\bibinfo {author} {\bibfnamefont {A.}~\bibnamefont
  {{Bolens}}}, \bibinfo {author} {\bibfnamefont {H.}~\bibnamefont {{Katsura}}},
  \bibinfo {author} {\bibfnamefont {M.}~\bibnamefont {{Ogata}}}, \ and\
  \bibinfo {author} {\bibfnamefont {S.}~\bibnamefont {{Miyashita}}},\
  }\href@noop {} {\bibfield  {journal} {\bibinfo  {journal} {ArXiv e-prints}\ }
  (\bibinfo {year} {2017})},\ \Eprint {http://arxiv.org/abs/1704.03153}
  {arXiv:1704.03153 [cond-mat.str-el]} \BibitemShut {NoStop}%
\bibitem [{\citenamefont {Bulaevskii}\ \emph {et~al.}(2008)\citenamefont
  {Bulaevskii}, \citenamefont {Batista}, \citenamefont {Mostovoy},\ and\
  \citenamefont {Khomskii}}]{Bulaevskii2008}%
  \BibitemOpen
  \bibfield  {author} {\bibinfo {author} {\bibfnamefont {L.~N.}\ \bibnamefont
  {Bulaevskii}}, \bibinfo {author} {\bibfnamefont {C.~D.}\ \bibnamefont
  {Batista}}, \bibinfo {author} {\bibfnamefont {M.~V.}\ \bibnamefont
  {Mostovoy}}, \ and\ \bibinfo {author} {\bibfnamefont {D.~I.}\ \bibnamefont
  {Khomskii}},\ }\href {\doibase 10.1103/PhysRevB.78.024402} {\bibfield
  {journal} {\bibinfo  {journal} {Phys. Rev. B}\ }\textbf {\bibinfo {volume}
  {78}},\ \bibinfo {pages} {024402} (\bibinfo {year} {2008})}\BibitemShut
  {NoStop}%
\bibitem [{\citenamefont {Potter}\ \emph {et~al.}(2013)\citenamefont {Potter},
  \citenamefont {Senthil},\ and\ \citenamefont {Lee}}]{Potter2013}%
  \BibitemOpen
  \bibfield  {author} {\bibinfo {author} {\bibfnamefont {A.~C.}\ \bibnamefont
  {Potter}}, \bibinfo {author} {\bibfnamefont {T.}~\bibnamefont {Senthil}}, \
  and\ \bibinfo {author} {\bibfnamefont {P.~A.}\ \bibnamefont {Lee}},\ }\href
  {\doibase 10.1103/PhysRevB.87.245106} {\bibfield  {journal} {\bibinfo
  {journal} {Phys. Rev. B}\ }\textbf {\bibinfo {volume} {87}},\ \bibinfo
  {pages} {245106} (\bibinfo {year} {2013})}\BibitemShut {NoStop}%
\bibitem [{\citenamefont {Pilon}\ \emph {et~al.}(2013)\citenamefont {Pilon},
  \citenamefont {Lui}, \citenamefont {Han}, \citenamefont {Shrekenhamer},
  \citenamefont {Frenzel}, \citenamefont {Padilla}, \citenamefont {Lee},\ and\
  \citenamefont {Gedik}}]{Pilon2013}%
  \BibitemOpen
  \bibfield  {author} {\bibinfo {author} {\bibfnamefont {D.~V.}\ \bibnamefont
  {Pilon}}, \bibinfo {author} {\bibfnamefont {C.~H.}\ \bibnamefont {Lui}},
  \bibinfo {author} {\bibfnamefont {T.~H.}\ \bibnamefont {Han}}, \bibinfo
  {author} {\bibfnamefont {D.}~\bibnamefont {Shrekenhamer}}, \bibinfo {author}
  {\bibfnamefont {A.~J.}\ \bibnamefont {Frenzel}}, \bibinfo {author}
  {\bibfnamefont {W.~J.}\ \bibnamefont {Padilla}}, \bibinfo {author}
  {\bibfnamefont {Y.~S.}\ \bibnamefont {Lee}}, \ and\ \bibinfo {author}
  {\bibfnamefont {N.}~\bibnamefont {Gedik}},\ }\href {\doibase
  10.1103/PhysRevLett.111.127401} {\bibfield  {journal} {\bibinfo  {journal}
  {Phys. Rev. Lett.}\ }\textbf {\bibinfo {volume} {111}},\ \bibinfo {pages}
  {127401} (\bibinfo {year} {2013})}\BibitemShut {NoStop}%
\bibitem [{\citenamefont {{Laurita}}\ \emph {et~al.}(2017)\citenamefont
  {{Laurita}}, \citenamefont {{Marcus}}, \citenamefont {{Trump}}, \citenamefont
  {{Kindervater}}, \citenamefont {{Stone}}, \citenamefont {{McQueen}},
  \citenamefont {{Broholm}},\ and\ \citenamefont {{Armitage}}}]{Laurita2017}%
  \BibitemOpen
  \bibfield  {author} {\bibinfo {author} {\bibfnamefont {N.~J.}\ \bibnamefont
  {{Laurita}}}, \bibinfo {author} {\bibfnamefont {G.~G.}\ \bibnamefont
  {{Marcus}}}, \bibinfo {author} {\bibfnamefont {B.~A.}\ \bibnamefont
  {{Trump}}}, \bibinfo {author} {\bibfnamefont {J.}~\bibnamefont
  {{Kindervater}}}, \bibinfo {author} {\bibfnamefont {M.~B.}\ \bibnamefont
  {{Stone}}}, \bibinfo {author} {\bibfnamefont {T.~M.}\ \bibnamefont
  {{McQueen}}}, \bibinfo {author} {\bibfnamefont {C.~L.}\ \bibnamefont
  {{Broholm}}}, \ and\ \bibinfo {author} {\bibfnamefont {N.~P.}\ \bibnamefont
  {{Armitage}}},\ }\href@noop {} {\bibfield  {journal} {\bibinfo  {journal}
  {ArXiv e-prints}\ } (\bibinfo {year} {2017})},\ \Eprint
  {http://arxiv.org/abs/1704.04228} {arXiv:1704.04228 [cond-mat.str-el]}
  \BibitemShut {NoStop}%
\bibitem [{\citenamefont {{Little}}\ \emph {et~al.}(2017)\citenamefont
  {{Little}}, \citenamefont {{Wu}}, \citenamefont {{Lampen-Kelley}},
  \citenamefont {{Banerjee}}, \citenamefont {{Pantankar}}, \citenamefont
  {{Rees}}, \citenamefont {{Bridges}}, \citenamefont {{Yan}}, \citenamefont
  {{Mandrus}}, \citenamefont {{Nagler}},\ and\ \citenamefont
  {{Orenstein}}}]{Little2017}%
  \BibitemOpen
  \bibfield  {author} {\bibinfo {author} {\bibfnamefont {A.}~\bibnamefont
  {{Little}}}, \bibinfo {author} {\bibfnamefont {L.}~\bibnamefont {{Wu}}},
  \bibinfo {author} {\bibfnamefont {P.}~\bibnamefont {{Lampen-Kelley}}},
  \bibinfo {author} {\bibfnamefont {A.}~\bibnamefont {{Banerjee}}}, \bibinfo
  {author} {\bibfnamefont {S.}~\bibnamefont {{Pantankar}}}, \bibinfo {author}
  {\bibfnamefont {D.}~\bibnamefont {{Rees}}}, \bibinfo {author} {\bibfnamefont
  {C.~A.}\ \bibnamefont {{Bridges}}}, \bibinfo {author} {\bibfnamefont {J.-Q.}\
  \bibnamefont {{Yan}}}, \bibinfo {author} {\bibfnamefont {D.}~\bibnamefont
  {{Mandrus}}}, \bibinfo {author} {\bibfnamefont {S.~E.}\ \bibnamefont
  {{Nagler}}}, \ and\ \bibinfo {author} {\bibfnamefont {J.}~\bibnamefont
  {{Orenstein}}},\ }\href@noop {} {\bibfield  {journal} {\bibinfo  {journal}
  {ArXiv e-prints}\ } (\bibinfo {year} {2017})},\ \Eprint
  {http://arxiv.org/abs/1704.07357} {arXiv:1704.07357 [cond-mat.str-el]}
  \BibitemShut {NoStop}%
\bibitem [{\citenamefont {Caux}\ and\ \citenamefont
  {Hagemans}(2006)}]{Caux2006}%
  \BibitemOpen
  \bibfield  {author} {\bibinfo {author} {\bibfnamefont {J.-S.}\ \bibnamefont
  {Caux}}\ and\ \bibinfo {author} {\bibfnamefont {R.}~\bibnamefont
  {Hagemans}},\ }\href {http://stacks.iop.org/1742-5468/2006/i=12/a=P12013}
  {\bibfield  {journal} {\bibinfo  {journal} {Journal of Statistical Mechanics:
  Theory and Experiment}\ }\textbf {\bibinfo {volume} {2006}},\ \bibinfo
  {pages} {P12013} (\bibinfo {year} {2006})}\BibitemShut {NoStop}%
\bibitem [{\citenamefont {Wang}\ and\ \citenamefont
  {Vishwanath}(2006)}]{Wang2006}%
  \BibitemOpen
  \bibfield  {author} {\bibinfo {author} {\bibfnamefont {F.}~\bibnamefont
  {Wang}}\ and\ \bibinfo {author} {\bibfnamefont {A.}~\bibnamefont
  {Vishwanath}},\ }\href {\doibase 10.1103/PhysRevB.74.174423} {\bibfield
  {journal} {\bibinfo  {journal} {Phys. Rev. B}\ }\textbf {\bibinfo {volume}
  {74}},\ \bibinfo {pages} {174423} (\bibinfo {year} {2006})}\BibitemShut
  {NoStop}%
\bibitem [{\citenamefont {Messio}\ \emph {et~al.}(2013)\citenamefont {Messio},
  \citenamefont {Lhuillier},\ and\ \citenamefont {Misguich}}]{Messio2013}%
  \BibitemOpen
  \bibfield  {author} {\bibinfo {author} {\bibfnamefont {L.}~\bibnamefont
  {Messio}}, \bibinfo {author} {\bibfnamefont {C.}~\bibnamefont {Lhuillier}}, \
  and\ \bibinfo {author} {\bibfnamefont {G.}~\bibnamefont {Misguich}},\ }\href
  {\doibase 10.1103/PhysRevB.87.125127} {\bibfield  {journal} {\bibinfo
  {journal} {Phys. Rev. B}\ }\textbf {\bibinfo {volume} {87}},\ \bibinfo
  {pages} {125127} (\bibinfo {year} {2013})}\BibitemShut {NoStop}%
\bibitem [{\citenamefont {{Iaconis}}\ \emph {et~al.}(2017)\citenamefont
  {{Iaconis}}, \citenamefont {{Liu}}, \citenamefont {{Hal{\'a}sz}},\ and\
  \citenamefont {{Balents}}}]{Iaconis2017}%
  \BibitemOpen
  \bibfield  {author} {\bibinfo {author} {\bibfnamefont {J.}~\bibnamefont
  {{Iaconis}}}, \bibinfo {author} {\bibfnamefont {C.}~\bibnamefont {{Liu}}},
  \bibinfo {author} {\bibfnamefont {G.~B.}\ \bibnamefont {{Hal{\'a}sz}}}, \
  and\ \bibinfo {author} {\bibfnamefont {L.}~\bibnamefont {{Balents}}},\
  }\href@noop {} {\bibfield  {journal} {\bibinfo  {journal} {ArXiv e-prints}\ }
  (\bibinfo {year} {2017})},\ \Eprint {http://arxiv.org/abs/1708.07856}
  {arXiv:1708.07856 [cond-mat.str-el]} \BibitemShut {NoStop}%
\end{thebibliography}%

\begin{widetext}
\renewcommand{\theequation}{S-\arabic{equation}}
  \setcounter{equation}{0}  
  \renewcommand{\thefigure}{S-\arabic{figure}}
   \setcounter{figure}{0}
\newpage
\begin{center}
\bf{Supplementary Information for
`` Spinon magnetic resonance of quantum spin liquids'' \\ Zhu-Xi Luo, Ethan Lake, Jia-Wei Mei, and Oleg A. Starykh}
\end{center}

\newcommand{\T}{\mathcal{T}}
\renewcommand{\r}{\mathbf{r}}
\newcommand{\C}{\bar{C}_6}

The supplementary material is arranged as follows: Part A discusses geometry of the lattice, PSG analysis is reviewed in Part B, spectra for U1A11 and U1A01 are presented in Part C, and the discussion of dynamical susceptibility can be found in Part D.

\section{A. The Lattice}

This section introduces the coordinate systems in both real space and Fourier space.

\subsection{A1. Real Space}
The rare-earth Yb atoms in YbMgGaO$_4$ form a triangular lattice. We use the following intralayer coordinates $\mathbf{r}=x \mathbf{a}_1+ y \mathbf{a}_2$, with
\begin{equation}
\mathbf{a}_1=(1,0),~~\mathbf{a}_2=\left(-\frac{1}{2},\frac{\sqrt{3}}{2}\right). 
\end{equation}

The intralayer symmetries are generated by $\{T_1,T_2,C_2,C_3\}$, where $T_1: (x,y)\rightarrow (x+1,y)$, $T_2: (x,y)\rightarrow (x,y+1)$ are translations along the $\mathbf{a}_1$ and $\mathbf{a}_2$ directions, $C_2: (x,y)\rightarrow (y,x)$ a two-fold rotation, $C_3:(x,y)\rightarrow (-y,x-y)$ a three-fold counterclockwise rotation and $I:(x,y)\rightarrow (-x,-y)$ is an inversion. This symmetry group is equivalent to another group generated by $\{T_1, T_2, C_2, \C\}$, where $\C\equiv C_3^{-1} I$ acts as $\C: (x,y)\rightarrow (x-y,x).$ We sketch the symmetry operations in Fig.\ref{fig:lattice}.

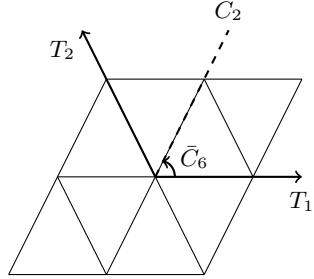
\begin{figure}[h]
	\centering
	\begin{tikzpicture}[scale=1.3]
	\draw (0,0)--(2,0);
	\draw (0.5,1)--(2.5,1);
	\draw (1,2)--(3,2);
	\draw (0,0)--(1,2);
	\draw (1,0)--(2,2);
	\draw (2,0)--(3,2);
	\draw (1,2)--(2,0);
	\draw (1,0)--(0.5,1);
	\draw (2.5,1)--(2,2);
	\draw[thick,->] (1.5,1)--(3,1);
	\draw[thick,->] (1.5,1)--(0.75,2.5);
	\draw[dashed,thick] (1.5,1)--(2.25,2.5);
	\draw[->,thick]  (1.7,1) to [out=90,in=-10] (1.57,1.16);
	\node at (2.25,2.7) {$C_2$};
	\node at (3,0.75) {$T_1$};
	\node at (0.55,2.3) {$T_2$};
	\node at (1.9,1.2) {$\C$};
	\end{tikzpicture}
	\caption{The symmetry operations.}
	\label{fig:lattice}
\end{figure}

These symmetry operations satisfy the following identities:
\begin{equation}
\label{eq:symmetry}
\begin{split}
& T_1^{-1}T_2T_1T_2^{-1}=T_1^{-1}T_2^{-1}T_1T_2=1,\\
& C_2^{-1}T_1C_2T_2^{-1}=C_2^{-1}T_2C_2T_1^{-1}=1,\\
& \C^{-1}T_1\C T_2=\C^{-1}T_2\C T_2^{-1}T_1^{-1}=1,\\
& (C_2)^2=(\C)^6=(\C C_2)^2=1.
\end{split}
\end{equation}

We further require a time reversal symmetry $\T^2=1$, satisfying
\begin{equation}
T_1^{-1}\T T_1\T=T_2^{-1}\T T_2 \T=1,~~C_2^{-1}\T C_2 \T=\C^{-1}\T \C \T=1.
\end{equation}

\subsection{A2. k-space}

The reciprocal lattice (Fig.\ref{fig:reciprocal}) $\mathbf{k}=k_1\mathbf{b}_1+k_2\mathbf{b}_2$ is studied in the dual basis $\mathbf{b}_1=2\pi (1,\frac{\sqrt{3}}{3})$ and $\mathbf{b}_2=2\pi (0,\frac{2\sqrt{3}}{3})$. Here $k_1, k_2\in [0, 2\pi)$. 

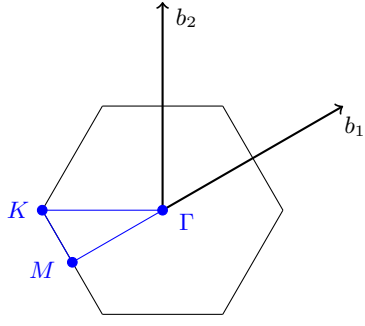
\begin{figure}[htbp]
	\centering
	\begin{tikzpicture}[scale=1.6]
	\draw (1,0)--(0.5,-0.866);
	\draw (0.5,-0.866)--(-0.5,-0.866);
	\draw (-0.5,-0.866)--(-1,0);
	\draw (-1,0)--(-0.5,0.866);
	\draw (-0.5,0.866)--(0.5,0.866);
	\draw (0.5,0.866)--(1,0);
	\draw[thick,->] (0,0)--(1.5,0.866);
	\draw[thick,->] (0,0)--(0,1.732);
	\node at (1.6,0.7) {$b_1$};
	\node at (0.2,1.6) {$b_2$};
	\draw[blue] (0,0)--(-1,0);
	\draw[blue] (0,0)--(-1.5/2,-0.866/2);
	\draw[blue] (-1,0)--(-1.5/2,-0.866/2);
	\filldraw[blue] (-1,0) circle [radius=0.04];
	\filldraw[blue] (0,0) circle [radius=0.04];
	\filldraw[blue] (-1.5/2,-0.866/2) circle [radius=0.04];
	\node at (0.2,-0.1) {\textcolor{blue}{$\Gamma$}};
	\node at (-1.2,0) {\textcolor{blue}{$K$}};
	\node at (-1,-0.5) {\textcolor{blue}{$M$}};
	\end{tikzpicture}
	\caption{The reciprocal lattice.}
	\label{fig:reciprocal}
\end{figure}

\section{B. PSG Analysis}
We review the projective symmetry group (PSG) analysis in this section, following the discussion in \cite{Li2016}. PSG is a powerful tool in the classification of symmetric spin liquids, which was first introduced in Ref.\cite{Wen2002} and lists all possible classes of lattice symmetry representations in the extended Hilbert space of spinons.

In the Nambu spinor representation $\Psi_{\r}=(f_{\r\uparrow},f_{\r\downarrow}^\dagger,f_{\r\downarrow},-f_{\r\uparrow}^\dagger)^T$, the mean field Hamiltonian is
\begin{equation}
\label{eq:HMF}
H_{MF}=-\frac{1}{2}\sum_{\r,\r'}\left(\Psi_{\r}^\dagger u_{\r\r'}\Psi_{\r'}+h.c.\right),
\end{equation} where $u$ is the hopping matrix. The PSG method, in a word, seeks to classify gauge-inequivalent mean-field ansatz $u_{\r\r'}$. In this Nambu representation, the spin operator $\mathbf{S}_\r$ and the generator $\mathbf{G}_\r$ of the $SU(2)$ gauge transformation are given by, \begin{equation}
\label{eq:SG}
\mathbf{S}_\r=\frac{1}{4}\Psi_\r^\dagger({\bm \sigma}\otimes \mathbf{1}_{2\times 2})\Psi_\r,~~\mathbf{G}_\r=\frac{1}{4}\Psi_\r^\dagger(\mathbf{1}_{2\times 2}\otimes {\bm \sigma})\Psi_\r.
\end{equation}

Under a symmetry operation $O$, $\Psi_\r$ transforms as
\begin{equation}
\Psi_\r\rightarrow \mathcal{U}_O\mathcal{G}^O_{O(\r)}\Psi_{O(\r)}=\mathcal{G}^O_{O(\r)}\mathcal{U}_O\Psi_{O(\r)},
\end{equation}
where $G^O_{O(\r)}$ is the local gauge transformation and $\mathcal{U}_O$ accounts for the rotation of the spin components due to the symmetry operation $O$.

The commutation in the above equation can be understood from the observation $[S_\r^\mu,G_\r^\nu]=0$. From equation \eqref{eq:SG}, one further convinces himself of the block diagonal form of the gauge transformation $G_\r^O=\mathbf{1}_{2\times 2}\otimes W_\r^O$, where $W_\r^O$ is a $2\times 2$-dimensional matrix.

The mean field ansatz  \eqref{eq:HMF} is invariant under the invariant gauge group ($IGG=U(1)$) with $u_{\r\r'}=\mathcal{G}_\r^{1\dagger}u_{\r\r'}\mathcal{G}_{O(\r')}^O$.
To respect a lattice symmetry transformation $O$,  the Hamiltonian should satisfy \cite{Reuther2014}
\begin{equation}
u_{\r\r'}=G_{O(\r)}^{O\dagger}\mathcal{U}_O^\dagger u_{O(\r)O(\r')}\mathcal{U}_O\mathcal{G}_{O(\r')}^O.
\end{equation}
In the isotropic case with full $SU(2)$ rotational symmetry, this matrix is trivial, i.e. $\mathcal{U}=\mathbf{1}_{2\times 2}$. For YbMgGaO$_4$, this symmetry is partially broken and reduced to $U(1)$, leading to a nontrivial $\mathcal{U}$. (Note that the spin rotation symmetry $U(1)$  is independent of the $U(1)$ gauge symmetry, it is the latter that gives the name of $U(1)$ spin liquid.)
 
A general group relation with the form $O_1O_2O_3O_4=1$ thus means the following constraint for the ansatz:
\begin{equation}
\mathcal{U}_{O_1}\mathcal{G}_\r^{O_1}\mathcal{U}_{O_2}\mathcal{G}_{O_2O_3O_4(\r)}^{O_2}\mathcal{U}_{O_3}\mathcal{G}_{O_3O_4(\r)}^{O_3}\mathcal{U}_{O_4}\mathcal{G}_{O_4(\r)}^{O_4}\in IGG.
\end{equation}

Using the commutation relation, we can rewrite it as \begin{equation}
\mathcal{U}_{O_1}\mathcal{U}_{O_2}\mathcal{U}_{O_3}\mathcal{U}_{O_4}\mathcal{G}_\r^{O_1}\mathcal{G}_{O_2O_3O_4(\r)}^{O_2}\mathcal{G}_{O_3O_4(\r)}^{O_3}\mathcal{G}_{O_4(\r)}^{O_4}\in IGG.
\end{equation}

As the series of rotations $O_1O_2O_3O_4$ either rotate the spinons by $0$ or $2\pi$, $\mathcal{U}_{O_1}\mathcal{U}_{O_2}\mathcal{U}_{O_3}\mathcal{U}_{O_4}=\pm \mathbf{1}_{4\times 4}$ always belongs to $IGG$. Thus the constraint further reduces to 
\begin{equation}
\mathcal{G}_\r^{O_1}\mathcal{G}_{O_2O_3O_4(\r)}^{O_2}\mathcal{G}_{O_3O_4(\r)}^{O_3}\mathcal{G}_{O_4(\r)}^{O_4}\in IGG,
\end{equation}
or equivalently,
\begin{equation}
\label{eq:constraint}
W_\r^{O_1}W_{O_2O_3O_4(\r)}^{O_2}W_{O_3O_4(\r)}^{O_3}W_{O_4(\r)}^{O_4}\in \{e^{i\phi\sigma_z}\mid\phi\in[0,2\pi)\}.
\end{equation}

For the space group symmetry of YbMgGaO$_4$, the nontrivial spin rotation matrices are
\begin{equation}
\label{eq:spin}
\mathcal{U}_{C_2}= \exp{(-i \frac{\pi}{2} \hat{n}_2\cdot {\bm \sigma})}=\left(\begin{matrix} & -e^{i\pi/6} \\ e^{-i\pi/6} &  \end{matrix}\right),~~~~
\mathcal{U}_{\C}= \exp{(i \frac{\pi}{3} \hat{n}_3\cdot {\bm \sigma})}=\left(\begin{matrix} e^{i\pi/3} &  \\  & e^{-i\pi/3} \end{matrix}\right),
\end{equation}
with $\hat{n}_2=\left(\frac{1}{2},\frac{\sqrt{3}}{2},0\right)$ is the rotation axis for the $C_2$ action and $\hat{n}_3=(0,0,1)$, since $\C\equiv C_3^{-1}I$ consists of a clockwise rotation with respect to the z-axis by $2\pi/3$.

For the gauge part, we choose the canonical gauge \cite{Wen2002} with $IGG=\{\mathbf{1}_{2\times 2}\otimes e^{i\phi\sigma_z}\mid \phi\in [0, 2\pi)\}$. Then the gauge transformation associated with the symmetry operation $O$ takes the form 
\begin{equation}
W_\r^O=(i\sigma_x)^{n_O}e^{i\phi_O[\r]\sigma_z},
\end{equation}
where $n_O\in\{0,1\}$.

In the case of translations, one can always choose a gauge so that
\begin{equation}
W_\r^{T_1}=(i\sigma_x)^{n_{T_1}},~~W_\r^{T_2}=(i\sigma_x)^{n_{T_2}}e^{i\phi_{T_2}[x,y]\sigma_z}.
\end{equation}
The constraints \eqref{eq:symmetry} further demand $n_1=n_2=0$ and $\phi_{T_2}[x+1,y]-\phi_{T_2}[x,y]\equiv \phi_1$, $\phi_{T_2}[x+1,y+1]-\phi_{T_2}[x,y+1]\equiv \phi_2$, with $\phi_1$ and $\phi_2$ to be determined. Here $\phi_1$ has the physical meaning of the flux through each unit cell of the triangular lattice. Since it is always possible to choose a gauge such that $\phi_{T_2}[0,y]=0$, we have $\phi_{T_2}[x,y]=\phi_1x$, thus $\phi_2=\phi_1$. Consequently, we obtain $W_\r^{T_1}=1$, $W_\r^{T_2}=e^{ix\phi_1\sigma_z}$. 

For $\C$ with $W_\r^{\C}=(i\sigma_x)^{n_{\C}}e^{i\phi_{\C}[x,y]\sigma_z}$, \eqref{eq:constraint} leads to \begin{equation}
\chi_{\C}[\r]=\phi_1xy-\phi_3x-\phi_4y-\phi_1\frac{y(y-1)}{2},
\end{equation}
with $\phi_3,\phi_4$ undetermined. When $n_{\C}=0$, $\phi_1$ is not constrained, while for $n_{\C}=1$, we further rquire $\phi_1=0,\pi$.

For $C_2$ with $W_\r^{C_2}=(i\sigma_x)^{n_{C_2}}e^{i\phi_{C_2}[x,y]\sigma_z}$, \eqref{eq:constraint} leads to $\phi_1=0,\pi$ and $\phi_3+2\phi_4=0$ for both $(n_{C_2},n_{\C})=(0,0)$ and $(0,1)$, while for $(n_{C_2},n_{\C})=(1,0)$ and $(1,1)$, $\phi_3$=0. In all  these four cases,
\begin{equation}
\begin{cases}
& \phi_{C_2}=-xy\phi_1,\\
& \phi_{\C}=xy\phi_1-\phi_1\frac{y(y-1)}{2}.
\end{cases}
\end{equation}

\section{C. The Spectra}

In this part we present the spectra of mean-field Hamiltonians U1A11 and U1A01.

\subsection{C1. U1A11}

In the U1A11 case, following the above section, one has $\phi_1=0$, $(n_{C_2},n_{\C})=(1,1)$, and \begin{equation}
\label{eq:U1A11}
W_{T_1}=W_{T_2}=\mathbf{1}_{2\times 2},~~W_{C_2}=W_{\C}=W_{\T}=i\sigma_y.
\end{equation}


Combining \eqref{eq:spin} and \eqref{eq:U1A11}, we obtain the transformation rules in Tab.\ref{tab:U1A11}.

	\begin{table}[h]
		\centering
	\begin{tabular}{ |c|c| }
				\hline
		Symmetry & Transformation Rules\\
		\hline
		$T_1$   \rule{0pt}{4ex}      &$ \begin{cases} f_{(x,y)\uparrow}\rightarrow  f_{(x+1,y)\uparrow} \\ f_{(x,y)\downarrow}\rightarrow  f_{(x+1,y)\downarrow} \end{cases}   $ \rule{0pt}{4ex}\\ \hline 
		$T_2$   \rule{0pt}{4ex}     &$\begin{cases} f_{(x,y)\uparrow}\rightarrow  f_{(x,y+1)\uparrow} \\ f_{(x,y)\downarrow}\rightarrow  f_{(x,y+1)\downarrow} \end{cases}$ \\ \hline
		$C_2$   \rule{0pt}{5ex}     &  $ \begin{cases} f_{(x,y)\uparrow}\rightarrow e^{i \pi/6} f^\dagger_{(y,x)\uparrow} \\ f_{(x,y)\downarrow}\rightarrow e^{-i\pi/6} f^\dagger_{(y,x)\downarrow} \end{cases}$\\ \hline
		$\C$  \rule{0pt}{5ex}      & $ \begin{cases} f_{(x,y)\uparrow}\rightarrow e^{i \pi/3} f^\dagger_{(x-y,x)\downarrow} \\ f_{(x,y)\downarrow}\rightarrow- e^{-i\pi/3} f^\dagger_{(x-y,x)\uparrow} \end{cases}$ \\ \hline
		$\T$ \rule{0pt}{4ex}  & $\begin{cases} f_{(x,y)\uparrow}\rightarrow f_{(x,y)\downarrow} \\ f_{(x,y)\downarrow}\rightarrow - f_{(x,y)\uparrow} \end{cases}$\\
		\hline
	\end{tabular}
    \caption{U1A11 PSG analysis.}
    \label{tab:U1A11}
    \end{table}

One can then find the Hamiltonian that is invariant under these symmetry operations: 
\begin{equation}
\begin{split}
H=\sum_{x,y} \Big\{
& t_1 \left[\right.
i f^\dagger_{(x+1,y)\uparrow} f_{(x,y)\uparrow}+i f^\dagger_{(x,y+1)\uparrow} f_{(x,y)\uparrow}- i f^\dagger_{(x+1,y+1)\uparrow} f_{(x,y)\uparrow} \\
&~~~~
- i f^\dagger_{(x+1,y)\downarrow} f_{(x,y)\downarrow} -i f^\dagger_{(x,y+1)\downarrow} f_{(x,y)\downarrow} +i f^\dagger_{(x+1,y+1)\downarrow} f_{(x,y)\downarrow}
\left.\right]\\ 
& +t_1'\left[\right.
e^{i\pi/3}f^\dagger_{(x+1,y)\uparrow} f_{(x,y)\downarrow}-f^\dagger_{(x,y+1)\uparrow} f_{(x,y)\downarrow}+e^{2i\pi/3}f^\dagger_{(x+1,y+1)\uparrow} f_{(x,y)\downarrow}\\
&~~~~~~~~ +e^{2i\pi/3} f^\dagger_{(x+1,y)\downarrow} f_{(x,y)\uparrow} +f^\dagger_{(x,y+1)\downarrow} f_{(x,y)\uparrow}+e^{i\pi/3}f^\dagger_{(x+1,y+1)\downarrow} f_{(x,y)\uparrow}
\left.\right]\\
& + t_2'\left[\right.
e^{i\pi/6}f^\dagger_{(x+1,y-1)\uparrow} f_{(x,y)\downarrow}+e^{5i\pi/6}f^\dagger_{(x+1,y+2)\uparrow} f_{(x,y)\downarrow}-if^\dagger_{(x-2,y-1)\uparrow} f_{(x,y)\downarrow}\\
&~~~~~~~~ +e^{5i\pi/6}f^\dagger_{(x+1,y-1)\downarrow} f_{(x,y)\uparrow}-i f^\dagger_{(x-2,y-1)\downarrow} f_{(x,y)\uparrow}+e^{i\pi/6}f^\dagger_{(x+1,y+2)\downarrow} f_{(x,y)\uparrow}
\left.\right] + h.c. \Big\}
\end{split}
\end{equation}
Here the $t_1$ and $t_1'$ terms describe the first-neighbor spin-preserving and spin-flipping hoppings, respectively. The $t_2'$ term denotes the second-neighbor spin-flipping hopping.
Also note that spin-preserving next-nearest hopping is absent, i.e., $t_2 =0$.

In $\mathbf{k}$-space, we have
\begin{equation}
\mathcal{H}(k_1,k_2)=\left(\begin{matrix} \epsilon_{\mathbf{k}} & \eta_{\mathbf{k}} \\ \eta_{\mathbf{k}}^* & - \epsilon_{\mathbf{k}}\end{matrix}\right),
\end{equation}
where the spin-preserving and spin-flipping terms are 
\begin{equation}
\epsilon_{\mathbf{k}}=t_1 \left(\sin k_1+\sin k_2 -\sin(k_1+k_2)\right),
\end{equation}
\begin{equation}
 \eta_{\mathbf{k}}=t_1'\left(e^{5i\pi/6}\sin k_1 -i \sin k_2 -e^{i\pi/6}\sin(k_1+k_2)\right)-t_2'\left(e^{2i\pi/3}\sin(k_1-k_2)-e^{i\pi/3}\sin(k_1+2k_2)-\sin(2k_1+k_2)\right).
\end{equation}
The resulting bands are $E_\nu (\mathbf{k}) = (-1)^\nu E(\mathbf{k})$, with $\nu\in\{1,2\}$ and 
\begin{equation}
E(\mathbf{k})= \sqrt{\epsilon_{\mathbf{k}}^2+|\eta_{\mathbf{k}}|^2}.
\end{equation}
The two bands touch at $\Gamma$ and $M$ points. The function $E(\mathbf{k})$ is plotted in Fig.\ref{fig:U1A11-spectra}.

\begin{figure}[htbp]
	\centering
	\includegraphics[scale=.55]{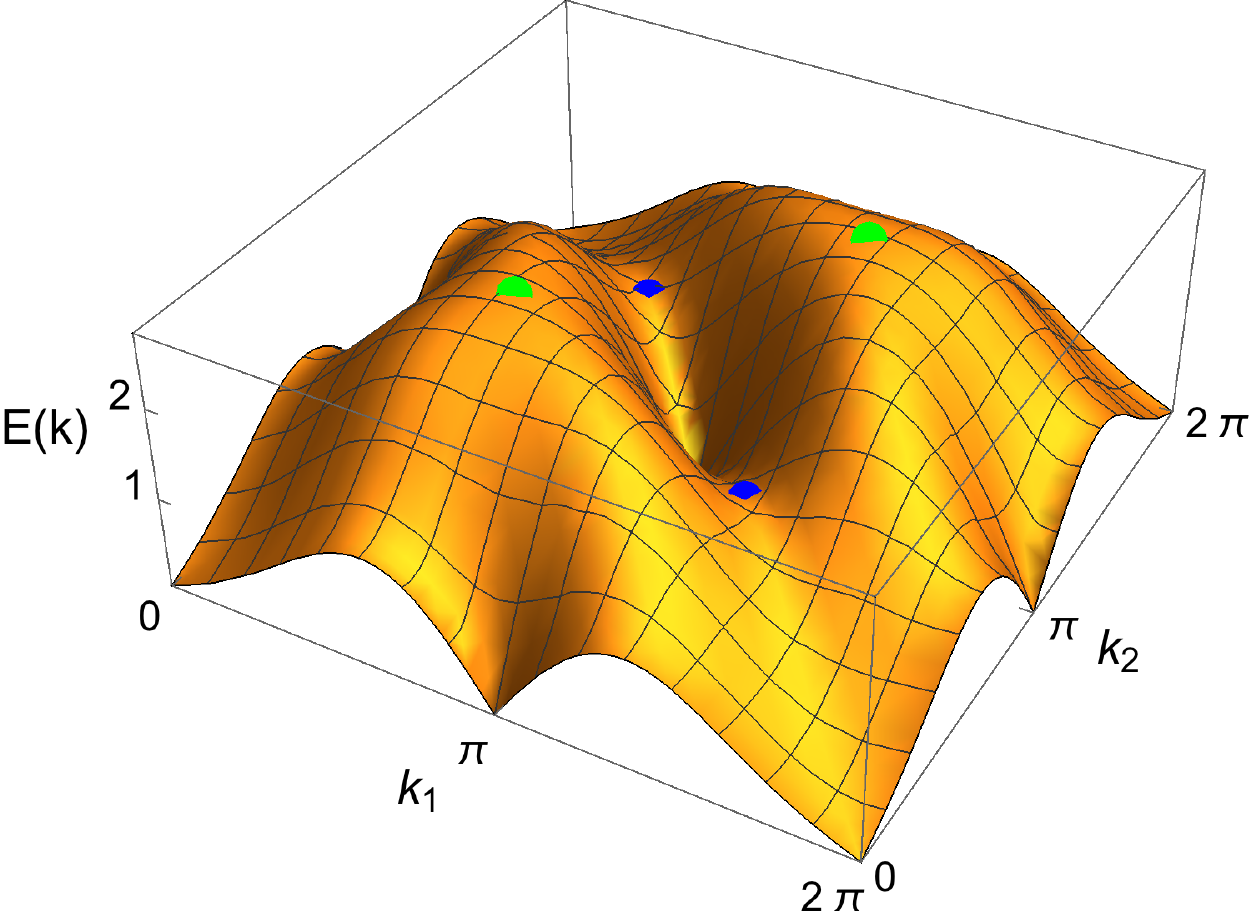}
	\caption{Spinon dispersion $E_2({\bf k})$ in U1A11 state (the upper band). Blue (green) dots indicate location of saddle points (maxima) which show up as singularities in $\chi''(\omega)$ in Fig.\ref{fig:1}. }
	\label{fig:U1A11-spectra}
\end{figure}

\subsection{C2. U1A01}

In the U1A01 case, $\phi_1=0$, $(n_{C_2},n_{\C})=(0,1)$, and the PSG analysis gives 
\begin{equation}
\label{eq:U1A01}
W_{T_1}=W_{T_2}=W_{C_2}=\mathbf{1}_{2\times 2},~~W_{\C}=W_{\T}=i\sigma_y.
\end{equation}

Combining with the spin rotation matrices, we obtain the following symmetry operations in Tab.\ref{tab:U1A01}.
	\begin{table}[h]
	\centering
	\begin{tabular}{ |c|c| }
		\hline
		Symmetry & Transformation Rules\\
		\hline
		$T_1$   \rule{0pt}{4ex}      &$ \begin{cases} f_{(x,y)\uparrow}\rightarrow  f_{(x+1,y)\uparrow} \\ f_{(x,y)\downarrow}\rightarrow  f_{(x+1,y)\downarrow} \end{cases}   $ \rule{0pt}{4ex}\\ \hline 
		$T_2$   \rule{0pt}{4ex}     &$\begin{cases} f_{(x,y)\uparrow}\rightarrow  f_{(x,y+1)\uparrow} \\ f_{(x,y)\downarrow}\rightarrow  f_{(x,y+1)\downarrow} \end{cases}$ \\ \hline
		$C_2$   \rule{0pt}{5ex}     &  $ \begin{cases} f_{(x,y)\uparrow}\rightarrow -e^{i \pi/6} f^\dagger_{(y,x)\downarrow} \\ f_{(x,y)\downarrow}\rightarrow e^{-i\pi/6} f^\dagger_{(y,x)\uparrow} \end{cases}$\\ \hline
		$\C$  \rule{0pt}{5ex}      & $ \begin{cases} f_{(x,y)\uparrow}\rightarrow e^{i \pi/3} f^\dagger_{(x-y,x)\downarrow} \\ f_{(x,y)\downarrow}\rightarrow- e^{-i\pi/3} f^\dagger_{(x-y,x)\uparrow} \end{cases}$ \\ \hline
		$\T$ \rule{0pt}{4ex}  & $\begin{cases} f_{(x,y)\uparrow}\rightarrow f_{(x,y)\downarrow} \\ f_{(x,y)\downarrow}\rightarrow - f_{(x,y)\uparrow} \end{cases}$\\
		\hline
	\end{tabular}
    \caption{U1A01 PSG analysis.}
    \label{tab:U1A01}
    \end{table}

The corresponding Hamiltonian invariant under such transformations is:
\begin{equation}
\begin{split}
H=\sum_{x,y} \Big\{
& t_1'\left[ e^{5i\pi/6}f^\dagger_{(x+1,y)\uparrow}-if^\dagger_{(x,y+1)\uparrow}+e^{-5i\pi/6}f^\dagger_{(x+1,y+1)\uparrow}\right]f_{(x,y)\downarrow}\\
& + t_1'\left[e^{i\pi/6}f^\dagger_{(x+1,y)\downarrow}-if^\dagger_{(x,y+1)\downarrow}+e^{-i\pi/6}f^\dagger_{(x+1,y+1)\downarrow}\right]f_{(x,y)\uparrow}\\
& + t_2'\left[e^{-i\pi/3}f^\dagger_{(x+1,y-1)\uparrow}+e^{i\pi/3} f^\dagger_{(x+1,y+2)\uparrow}+f^\dagger_{(x+2,y+1)\uparrow}\right]f_{(x,y)\downarrow}\\
&+ t_2' \left[e^{-2i\pi/3}f^\dagger_{(x+1,y-1)\downarrow}+e^{2\pi i/3}f^\dagger_{(x+1,y+2)\downarrow}-f^\dagger_{(x+2,y+1)\downarrow}\right]f_{(x,y)\uparrow}\\
&+ t_2 \left[ -i f^\dagger_{(x+1,y-1)\uparrow}-i f^\dagger_{(x+1,y+2)\uparrow}+if^\dagger_{(x+2,y+1)\uparrow}\right]f_{(x,y)\uparrow}\\
&+ t_2 \left[if^\dagger_{(x+1,y-1)\downarrow}+if^\dagger_{(x+1,y+2)\downarrow}-if^\dagger_{(x+2,y+1)\downarrow}\right]f_{(x,y)\uparrow}
 + h.c. \Big\},
\end{split}
\end{equation}
where again the $t_{1,2}$ describe the spin-preserving nearest-neighbor and next-nearest-neighbor hoppings and $t_{1,2}'$ describe the spin-flipping hoppings.
Note that now $t_1 = 0$, which most likely makes U1A01 state less energetically favorable than U1A11 one.
In $\mathbf{k}$-space, we again have
\begin{equation}
\mathcal{H}(k_1,k_2)=\left(\begin{matrix} \epsilon_{\mathbf{k}} & \eta_{\mathbf{k}} \\ \eta_{\mathbf{k}}^* & -\epsilon_{\mathbf{k}}\end{matrix}\right),
\end{equation}
now with
\begin{equation}
\begin{split}
\epsilon_{\mathbf{k}}= & t_2 \left[\sin(k_1-k_2)+\sin(k_1+2k_2)-\sin(2k_1+k_2)\right],\\-
\eta_{\mathbf{k}}= & t_1'\left[e^{-2i\pi/3}\sin k_1+\sin k_2+e^{-i\pi/3}\sin (k_1+k_2)\right] \\
& +t_2' \left[ e^{i\pi/6}\sin (k_1-k_2)+e^{5i\pi/6}\sin (k_1+2k_2)+i\sin(2k_1+k_2)\right].
\end{split}
\end{equation}

The spectra $E({\bf k})$ is plotted in Fig.\ref{fig:U1A01-spectra}. Spinon dispersions of the two U1A states along the path ${\bf b}_1 + {\bf b}_2$ in the reciprocal lattice, which contains $\Gamma$, K and M points, are compared in Figure \ref{fig:band}.

\begin{figure}[h]
	\centering
	\includegraphics[scale=.55]{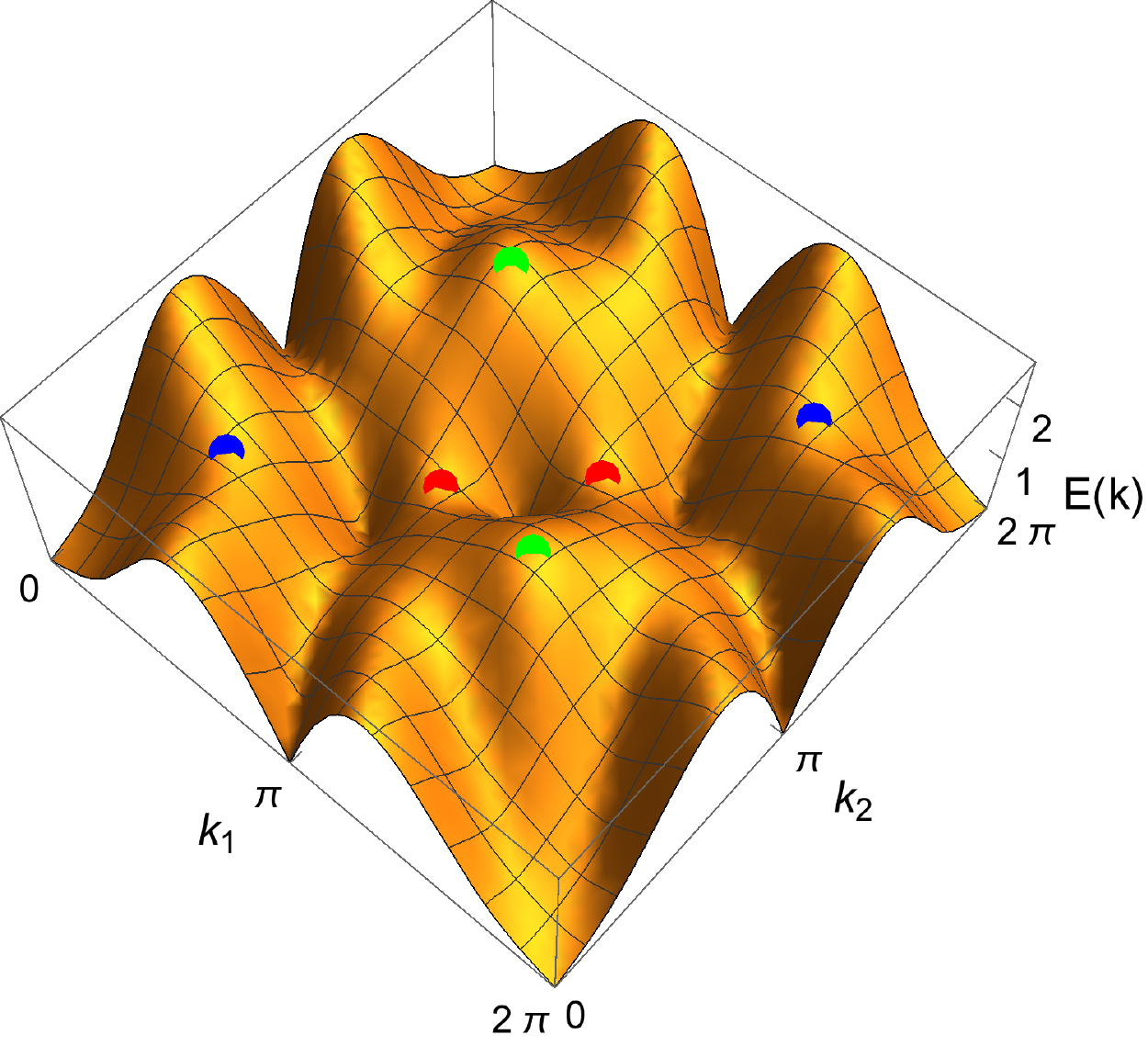}
	 \caption{Spinon dispersion $E_2({\bf k})$ in U1A01 state (the upper band). Blue and red (green) dots indicate location of saddle points (maxima) which show up as singularities of $\chi''(\omega)$ in Fig. \ref{fig:2}.
    The red saddle point, located between the K and M points of the Brillouin zone, is caused by the appearance of Dirac node at the K point in U1A01 state.}
	\label{fig:U1A01-spectra}
\end{figure}

\begin{figure}[h]
	\centering
    \includegraphics[scale=0.5]{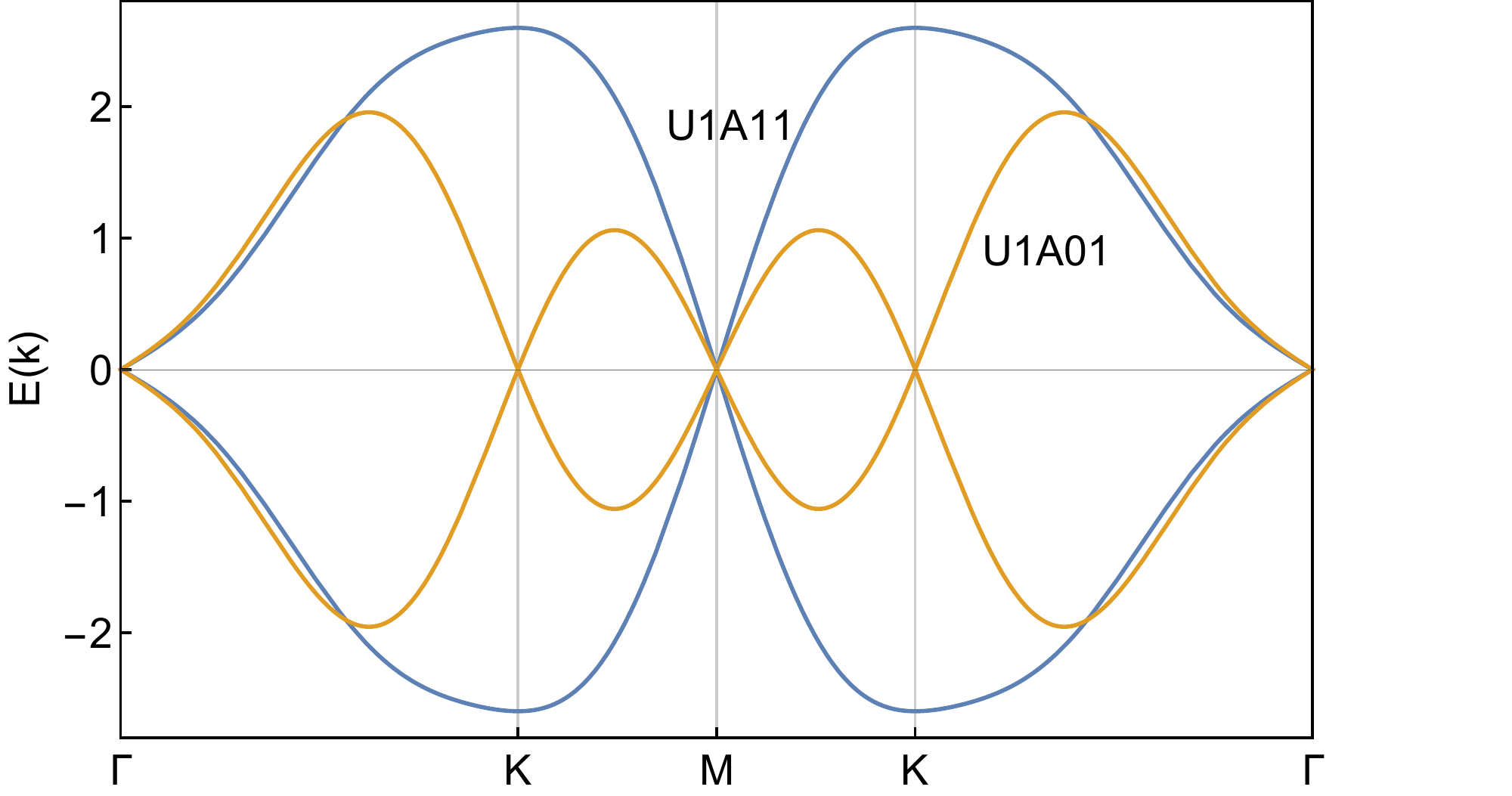}
    \caption{Spinon dispersions $E_{1,2}({\bf k})$ along the line $\Gamma$-K-M-K-$\Gamma$ for U1A11 and U1A01 states.}
    \label{fig:band}
\end{figure}

\subsection{C3. With Magnetic Field}
When an external magnetic field ${\bf B}=B_z\hat{\bf z}$ is present, the spectrum is shown in Fig.\ref{fig:U1A11-spectra-mag}.

\begin{figure}[h]
	\centering
	\includegraphics[scale=.55]{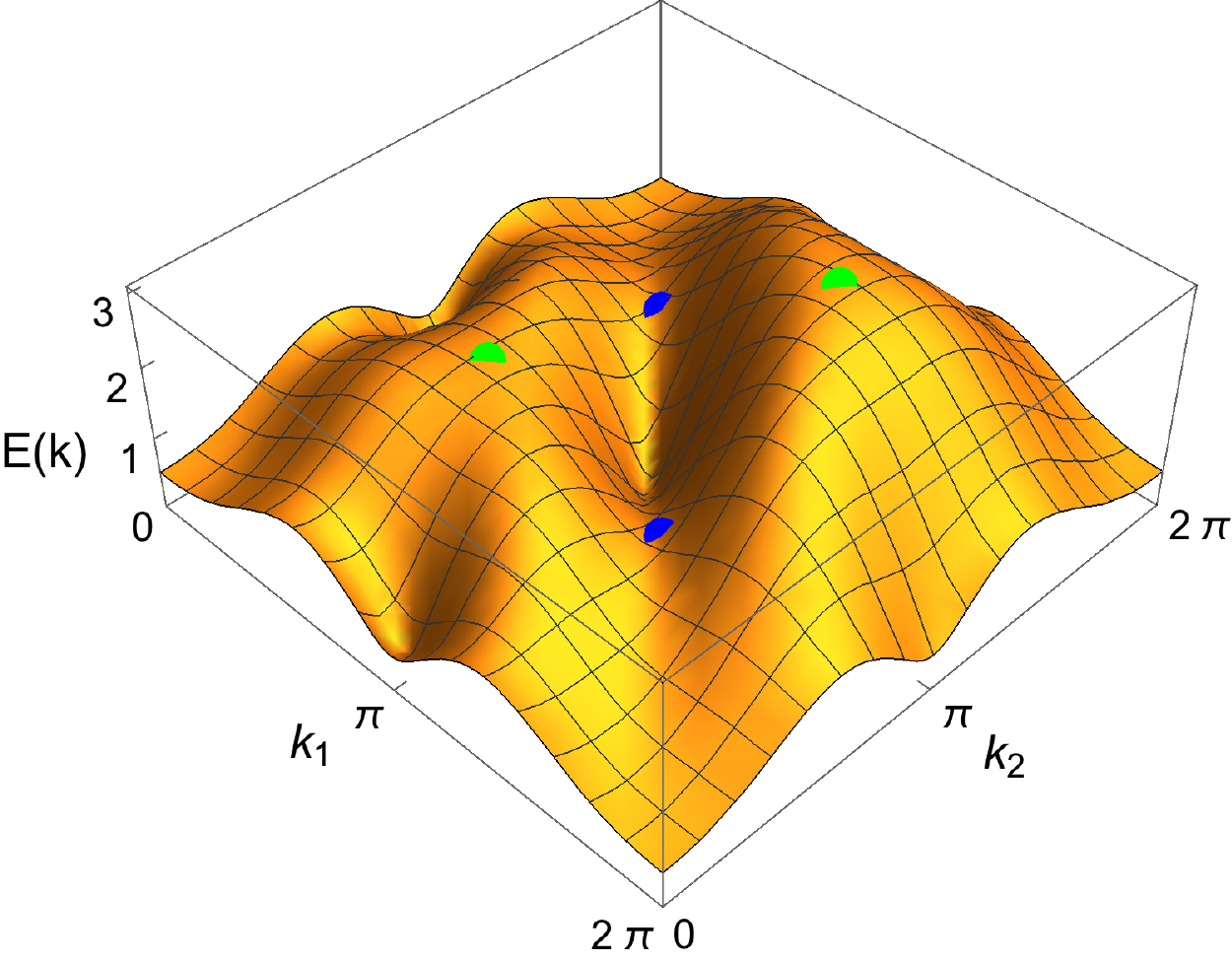}
	 \caption{Spinon dispersion $E_2({\bf k})$ in U1A11 state in the presence of magnetic field along $\hat{z}$, $B_z =1$. 
	 Blue (green) dots indicate location of saddle points (maxima) which show up as singularities in $\chi''(\omega)$ in Fig. \ref{fig:3}.
    Magnetic field along $\hat{z}$ breaks symmetry between the two maxima. The lower local maximum in the dispersion is causing discontinuity in $\chi''(\omega)$ at $\omega \approx 4.2$.}
	\label{fig:U1A11-spectra-mag}
\end{figure}

\clearpage

\section{D. The Dynamical Susceptibility}

Following equation \eqref{eq:5} in the text, the imaginary part of dynamical susceptibility is
\begin{eqnarray}
&&\chi_{nn}^{\prime\prime}(\omega)=-\frac{\pi}{4N}\sum_{\mathbf{k}}\frac{\delta (\omega-2E_1(\mathbf{k})}{E_1(\mathbf{k})^2}\Big[\left(\epsilon_{\mathbf{k}}^2+\eta_{\mathbf{k}}^{\prime\prime 2}\right)\sin^2\theta\cos^2\phi + \left(\epsilon_{\mathbf{k}}^2+\eta_{\mathbf{k}}^{\prime 2}\right)\sin^2\theta\sin^2\phi +|\eta_{\mathbf{k}}|^2  \cos^2\theta \nonumber\\
&&+2\eta'_{\bf k} \eta''_{\bf k}\sin\theta \cos\phi \sin\phi -2 \epsilon_{\bf k} \eta'_{\bf k} \sin\theta \cos\theta  \cos\phi + 2 \epsilon_{\bf k} \eta''_{\bf k}\sin\theta \cos\theta  \sin\phi  \Big], 
\end{eqnarray}
where $\eta^\prime_{\mathbf{k}}$ and $\eta^{\prime\prime}_{\mathbf{k}}$ denote the real and imaginary parts of $\eta_{\mathbf{k}}$, respectively. 
The 2nd line of this equation is made of off-diagonal terms which were not shown in the main text because their contribution vanishes identically.
This follows from the invariance of $\chi$ under symmetry transformations.

	\begin{table}[h]
		\centering
		\begin{tabular}{ |c|c|c|c| }
			\hline
			Symmetry & $\mathbf{a}_i$ & $\mathbf{b}_i$ & $k_i$\\
			\hline
			$C_2$   \rule{0pt}{5ex}      &$\begin{cases}  \mathbf{a}_1\rightarrow\mathbf{a}_2\\ \mathbf{a}_2\rightarrow\mathbf{a}_1  \end{cases}$ & $\begin{cases} \mathbf{b}_1\rightarrow\mathbf{b}_2\\ \mathbf{b}_2\rightarrow\mathbf{b}_1 \end{cases}$   \rule{0pt}{5ex} & $\begin{cases} k_1\rightarrow k_2\\ k_2\rightarrow k_1\end{cases}$ \\ \hline 
			$C_3$   \rule{0pt}{5ex}    &$\begin{cases}  \mathbf{a}_1\rightarrow\mathbf{a}_2\\ \mathbf{a}_2\rightarrow-\mathbf{a}_1-\mathbf{a}_2  \end{cases}$ & $\begin{cases} \mathbf{b}_1\rightarrow\mathbf{b}_2-\mathbf{b}_1\\ \mathbf{b}_2\rightarrow-\mathbf{b}_1\end{cases}$  &$\begin{cases} k_1\rightarrow-k_1-k_2 \\ k_2\rightarrow k_1\end{cases}$  \rule{0pt}{5ex}\\ \hline 
		\end{tabular}
		\caption{Transformation of coordinates under rotations.}
		\label{tab:off-diagonal}
	\end{table}

Under the two- and three-fold rotations, the bases in real and reciprocal space transform as in Tab.\ref{tab:off-diagonal}. $\chi$ transform differently for different mean-field Hamiltonians, due to their distinct dependences on ${\bf k}$.

\subsection{D1. U1A11}

For the U1A11 Hamiltonian, using explicit form of $\epsilon_{\bf k}$ and $\eta_{\mathbf{k}}$ we find that $\epsilon_{\mathbf{k}}$ is invariant while $\eta_{\mathbf{k}}$ changes as:
\begin{equation}
\begin{split}
C_2: &~\epsilon_{\mathbf{k}}\rightarrow \epsilon_{\mathbf{k}},~ ~\eta_{\mathbf{k}}^\prime\rightarrow \frac{1}{2} \eta_{\mathbf{k}}^\prime+\frac{\sqrt{3}}{2} \eta_{\mathbf{k}}^{\prime\prime},~~\eta_{\mathbf{k}}^{\prime\prime}\rightarrow \frac{\sqrt{3}}{2} \eta_{\mathbf{k}}^\prime-\frac{1}{2} \eta_{\mathbf{k}}^{\prime\prime},\\
C_3: &~\epsilon_{\mathbf{k}}\rightarrow \epsilon_{\mathbf{k}},~ ~\eta_{\mathbf{k}}^\prime\rightarrow -\frac{1}{2}\eta_{\mathbf{k}}^\prime-\frac{\sqrt{3}}{2}\eta_{\mathbf{k}}^{\prime\prime},~~\eta_{\mathbf{k}}^{\prime\prime}\rightarrow\frac{\sqrt{3}}{2}\eta_{\mathbf{k}}^\prime-\frac{1}{2}\eta_{\mathbf{k}}^{\prime\prime}.
\end{split}
\label{eq:sym1}
\end{equation}

We start from the analysis of $\hat{n}^x\hat{n}^z$ component of the susceptibility $\chi_{xz}^{\prime\prime}$, which is proportional to $\int d\mathbf{k} \epsilon_{\mathbf{k}} \eta_{\mathbf{k}}^\prime$.  Since the Brillouin zone is invariant under these $C_2$ and $C_3$ rotations, we only need to look at the integrand. From equation \eqref{eq:sym1}, one observes that $\epsilon_{\mathbf{k}}$ is invariant under both $C_2$ and $C_3$, while $\eta_{\mathbf{k}}^\prime$ obtains different signs under the two actions. Consequently, these two symmetries constrain $\chi_{xz}^{\prime\prime}$ to be zero.

We now proceed to look at the $\hat{n}^y\hat{n}^z$ component, which is proportional to $ \epsilon_{\mathbf{k}} \eta_{\mathbf{k}}^{\prime\prime}$. A $C_3$ rotation will lead to $\eta_{\mathbf{k}}^{\prime\prime}\rightarrow\frac{\sqrt{3}}{2}\eta_{\mathbf{k}}^\prime-\frac{1}{2}\eta_{\mathbf{k}}^{\prime\prime}$. But since the integration over $\epsilon_{\mathbf{k}} \eta_{\mathbf{k}}^{\prime}$ vanishes, the only value for 
$\int d\mathbf{k} ~ \epsilon_{\mathbf{k}} \eta^{\prime\prime}_{\bf k}$ that preserves the symmetry is zero. 

Finally, a similar argument leads to the vanishing of $\chi_{xy}^{\prime\prime} \propto \int d\mathbf{k}\eta_{\mathbf{k}}^\prime~ \eta_{\mathbf{k}}^{\prime\prime}$ as well.
These findings can be also verified by numerical integration of the involved expressions over the Brillouin zone. 

\subsection{D2. U1A01}
For U1A01 class, we instead have
\begin{equation}
\begin{split}
C_2: &~\epsilon_{\mathbf{k}}\rightarrow -\epsilon_{\mathbf{k}},~ ~\eta_{\mathbf{k}}^\prime\rightarrow -\frac{1}{2} \eta_{\mathbf{k}}^\prime-\frac{\sqrt{3}}{2} \eta_{\mathbf{k}}^{\prime\prime},~~\eta_{\mathbf{k}}^{\prime\prime}\rightarrow -\frac{\sqrt{3}}{2} \eta_{\mathbf{k}}^\prime+\frac{1}{2} \eta_{\mathbf{k}}^{\prime\prime},\\
C_3: &~\epsilon_{\mathbf{k}}\rightarrow \epsilon_{\mathbf{k}},~ ~\eta_{\mathbf{k}}^\prime\rightarrow -\frac{1}{2}\eta_{\mathbf{k}}^\prime-\frac{\sqrt{3}}{2}\eta_{\mathbf{k}}^{\prime\prime},~~\eta_{\mathbf{k}}^{\prime\prime}\rightarrow\frac{\sqrt{3}}{2}\eta_{\mathbf{k}}^\prime-\frac{1}{2}\eta_{\mathbf{k}}^{\prime\prime}.
\end{split}
\end{equation}
Same arguments lead to the vanishing of off-diagonal terms of the dynamical susceptibility.

\subsection{D3. Van Hove singularities}
The peak in the middle of the absorption band in Fig.1, at $\omega \approx 2.43$, is caused by the van Hove singularity, of the saddle point kind, of 
spinon dispersion $E_2({\bf k})$ at ${\bf k}_0 = (k_0, 2\pi - k_0)$ and also at $(2\pi-k_0, k_0)$. Here $k_0 \approx 1.97$
and $E_2({\bf k}\approx {\bf k}_0) \approx 1.215 + 1.31 (k_1 + k_2)^2 - 0.23 (k_1 - k_2)^2$. The saddle point region contributes to $\chi''$ as follows
(here $u = k_1 + k_2$, $v = k_1 - k_2$, we assume $\omega > 2E_2({\bf k}_0)$ and rescale $u, v$ to obtain unit coefficients in front of $u^2, v^2$ terms)
\begin{equation}
\chi''(\omega)_{\rm SP} \sim \int du \int dv~ \delta\left(\omega - 2E_2({\bf k}_0) - u^2 + v^2\right) = \int_{-\Lambda}^\Lambda \frac{d v}{\sqrt{\omega - 2E_2({\bf k}_0) + v^2}} = 
\ln\Big(\frac{4\Lambda^2}{\omega - 2E_2({\bf k}_0)}\Big)
\end{equation}
For $\omega < 2E_2({\bf k}_0)$ one first integrates over $v$, and then over $u$, to obtain the same result. We also used the fact that the prefactor of the delta-function in the expression for $\chi''$
reduces to a constant at ${\bf k}_0$. 
Therefore saddle point 
contributes $\chi''(\omega)_{\rm SP}  \sim \ln(\Lambda^2/|\omega - 2E_2({\bf k}_0)|)$.

To understand behavior near the upper edge of the continuum we need to consider spinon spectrum near its maximum. For U1A11 the maximum occurs near K point where 
$E_2({\bf k}) = E_{\rm max} - \beta (k_1^2 + k_1 k_2 + k_2^2)$, where $E_{\rm max} = 3\sqrt{3}/2$ and $\beta \approx 0.38$. We also need to know that near K point 
$|\eta_{\bf k}|^2 \to k_1^2 + k_1 k_2 + k_2^2$ while $\epsilon_{\bf k} \to \text{const}$. As a result $\chi''_{zz}$ (that is, $\theta=0$ part of susceptibility) and 
of $\chi''_{xx}$ (its $\theta = \pi/2$ part) behave differently near $\omega = 2 E_{\rm max}$.

Using $k_1 = r \cos\phi, k_2 = r \sin\phi$ we obtain after simple manipulations
\begin{equation}
\chi''_{zz}(\omega) \sim \int d\phi \int_0^\infty r^3 dr~ \delta\left(r^2 - \frac{2 E_{\rm max} - \omega}{\beta(1 + \cos\phi \sin\phi)}\right) \sim (2 E_{\rm max} - \omega) \Theta(2 E_{\rm max} - \omega)
\end{equation}
At the same time
\begin{equation}
\chi''_{xx}(\omega) \sim \int d\phi \int_0^\infty rdr~ \delta\left(r^2 - \frac{2 E_{\rm max} - \omega}{\beta(1 + \cos\phi \sin\phi)}\right) \sim \Theta(2 E_{\rm max} - \omega)
\end{equation}
reduces to a step-function. This explains behavior of $\omega \chi''$ for different polarizations near $\omega = 2 E_{\rm max} = 3\sqrt{3}$ in Fig.1.
Similar features are observed in Figures \ref{fig:2} and \ref{fig:3}. Figure \ref{fig:3} features contributions from two different local maxima that spinon dispersion develops under magnetic field along $\hat{z}$ axis,
see Fig.\ref{fig:U1A11-spectra-mag}. The smaller of these contributes to discontinuity of $\chi''$ at $\omega \approx 4.1$, while the global maximum of the spectrum controls 
behavior near the upper boundary, at $\omega \approx 6.2$ in Figure \ref{fig:3}.

\end{widetext}
\end{document}